\documentclass[12pt,preprint]{aastex}

\shorttitle{3D Structure of the Thick Disk}

\begin{document}

\title{The 3D Velocity Structure of the Thick Disk from SPM4 and RAVE DR2}

\author{Dana I. Casetti-Dinescu\altaffilmark{1,2}, 
Terrence M. Girard\altaffilmark{1}, 
Vladimir I. Korchagin\altaffilmark{3,1} and
William F. van Altena\altaffilmark{1}}

\altaffiltext{1}{Astronomy Department, Yale University, P.O. Box 208101,
New Haven, CT 06520-8101, USA, dana.casetti@yale.edu,terry.girard@yale.edu,william.vanaltena@yale.edu}
\altaffiltext{2}{Astronomical Institute of the Romanian Academy, Str.
Cutitul de Argint 5, RO-75212, Bucharest 28, Romania}
\altaffiltext{3}{Institute of Physics, Rostov University, Rostov-on-Don 344090, Russia, vkorchagin@sfedu.ru}

\begin{abstract}
We analyze the 3D kinematics of a sample of $\sim 4400$ red clump stars ranging
between 5 and 10 kpc from the Galactic center and
 up to 3 kpc from the Galactic plane. This sample is representative for the metal-rich 
([Fe/H] = -0.6 to 0.5) thick disk.
Absolute proper motions are from the fourth release of the Southern
Proper Motion Program, and radial velocities from the second release of the 
Radial Velocity Experiment. 
The derived kinematical properties of the thick disk include:
 the rotational velocity gradient 
$\partial V_{\theta} / \partial z = -25.2 \pm 2.1$ km~s$^{-1}$ kpc$^{-1}$, velocity dispersions  
$(\sigma_{V_R}, \sigma_{V_{\theta}}, \sigma_{V_z})\vert_{z=1} = (70.4, 48.0, 36.2) \pm(4.1,8.3,4.0)$ km~s$^{-1}$, and
velocity-ellipsoid tilt angle $\alpha_{Rz} = 8.6\arcdeg \pm 1.8 \arcdeg$.
Our dynamical estimate of the thin-disk 
scale length is $R_{thin} = 2.0 \pm 0.4$ kpc and the thick-disk
scale height is $z_{thick} = 0.7 \pm 0.1$ kpc.

The observed orbital eccentricity distribution 
compared with those from four different models of the formation of
the thick disk from Sales et al.~favor the gas-rich merger model and the minor merger heating model.

Interestingly, when referred to the currently accepted value of the LSR, stars more distant than 
0.7 kpc from the Sun show a net
average radial velocity of $13 \pm 3 $ km~s$^{-1}$.
This result is seen in previous kinematical studies
using other tracers at distances larger than $\sim 1$ kpc. We 
suggest this motion reflects an inward perturbation of the locally-defined LSR induced by
the spiral density wave.

\end{abstract}


\section{Introduction}

Recent large databases of absolute proper motions and radial velocities 
offer an entirely new perspective on the 3D kinematical structure of the Milky Way,
as they probe large volumes unavailable in previous studies.
Such recent studies are primarily based on the
Sloan Digital Sky Survey (SDSS, York et al. 2000)  and its sub-survey Sloan Extension for
Galactic Understanding and Exploration (SEGUE, Yanny et al. 2009).
For instance, Carollo et al. (2010, C10), Bond et al. (2010, B10), Smith et al. (2009)
have characterized in detail the inner and outer halo, and the disk (thin, thick and metal weak).
Proper motions in these studies are derived using positions from the USNO-B catalog (Monet et al. 2003) and 
SDSS (Munn et al. 2004, 2008),
or from SDSS-only data (Bramich et al. 2008). 

Another significant effort in this direction is the combination of the RAdial Velocity Experiment (RAVE,
e.g. for Data Release 2, Zwitter et al. 2008) with other proper-motion catalogs (e.g., UCAC2 - Zacharias 
et al. 2000, Tycho2 - Perryman et al. 1997), as exemplified in
studies by Siebert et al. (2008), Veltz et al. (2008). 

In this work we combine radial velocities from RAVE DR2 with absolute proper motions from the fourth release of the 
Southern Proper Motion Catalog (SPM4, Girard et al. 2010), to provide an accurate 3D kinematical description of the thick disk. 
It is  particularly desirable to have such a description in order to help discriminate among competing models
for the formation of this component.

The work by Sales et al. (2009) shows that the orbital eccentricity distribution is a good discriminator between
various scenarios such as; accretion of many satellites, radial migration, dynamical heating from a satellite, and
gas-rich merging with satellites and in-situ star formation. 
Two recent papers based on RAVE (Wilson et al. 2010) and SDSS data (Dierickx et al. 2010) use the observed eccentricity distribution to favor
the gas-rich merging scenario. However the SDSS-based study by Loebman et al. (2010) argues in favor of radial migration, primarily
based on the lack of correlation between rotation velocity and metallicity. 
Thus, a number of issues remain to be explored both on the modeling and observational side for
a realistic description of the formation of the thick disk. 

Also, the recent study by Bond et al. (2010) determines much lower velocity dispersions of the thick 
disk than traditionally measured (e.g., Soubiran et al. 2003, Chiba \& Beers 2000), raising yet again the question
of a clear distinction between the thin and thick disks. With these issues in mind, we hope this contribution will help
improve the current understanding of the thick disk.

Our sample consists of $\sim 4400$ red clump stars primarily selected using 2MASS photometry (Skrutskie et al. 2006). It is a clean sample of
well-behaved tracers for which both sources of radial velocities and proper motions are well understood in terms of
systematic uncertainties. The sample includes stars  between 0.4 and 4 kpc from the Sun, which corresponds to
a distance between 5 and 10 kpc from the GC, and up to 3 kpc from the Galactic plane.
This sample is used to determine various kinematical parameters of the thick disk,
a dynamical estimate of the thin-disk scale length to
be compared with star-count determinations and the eccentricity distribution of stars in the thick disk.
Our results are then discussed in the framework of current thick-disk formation models.

The paper is organized as follows: in Section 2 we describe the sample selection and its properties, in Section 3 we show the spatial distribution
of the sample and the calculation of velocities. Section 4 discusses the LSR velocity, while Section 5 details the velocity and
velocity dispersion gradients. In Section 6, we determine the tilt angles of the velocity ellipsoid, and in Section 7 we determine the
scale length of the thin disk. The comparison of the eccentricity distributions is presented in Section 8. 
In Section 9, we summarize our results.

\section{The Sample: Selection and Properties}

The RAVE DR2 dataset has 49327 individual objects, of which 21121 also have stellar parameters
such as surface gravity, effective temperature and metallicity measured.
Typical uncertainties for surface gravity, effective temperature and metallicity are
0.5 dex, 400 K and 0.2 dex respectively (Zwitter et al. 2008).

The SPM4 catalog contains positions and absolute proper motions for $\sim 103$ million objects
down to $V = 17.5$. Absolute proper motions are 
tied to the ICRS via $Hipparcos$ stars at the bright end,
and to galaxies at the faint end. The catalog also provides $BV$ photometry
from various sources including CCD photometry from the second-epoch SPM data and 2MASS $JHK$ photometry.
Individual proper-motion uncertainties vary with magnitude; well measured stars ($V \sim 12$) have 
uncertainties of $\sim 1.0$ mas~yr$^{-1}$. 
According to Girard et al. (2010) the proper-motion error estimates in SPM4 are reliable, based on a comparison with 
the more precise SPM2 catalog (Platais et al. 1998). 
The accuracy of the absolute proper-motion system
is expected to be of the order of $\sim 1.0$ mas~yr$^{-1}$ (Girard et al. 2010).

We match the SPM4 catalog with the entire RAVE DR2 dataset, to obtain 
31994 unique objects. The match is done by positional coincidence with a 
maximum matching radius of $1\arcsec$. For multiple matching, we choose the object with 
the smallest separation. 
In Figure 1, we show the distribution of RAVE and SPM4 data in equatorial 
(top), and galactic (bottom) coordinates.

For each object, we determine the reddening $E_{B-V}$ from the Schlegel et al. (1998) maps 
and eliminate from further analysis objects with $E_{B-V} > 0.5$. 
Objects are dereddened using the relationships from Majewski et al. (2003).
We will focus on the analysis of red clump stars, which are metal-rich stars in the He-burning core phase
with well defined absolute magnitudes, and thus distances. The absolute magnitude of red clump stars
is $M_K = -1.61$ with a dispersion of 0.22 mag (Alves 2000), and it varies little with 
metallicity [Fe/H] if the metallicity range is between -0.5 and 0.1 (Grocholski \& Sarajedini 2002).
In Figure 2 we show the $M_K, (J-K)_{0}$ HR diagram for $Hipparcos$ stars that have 2MASS photometry. These stars
are south of Dec $= -20\arcdeg$ as in the SPM4 area coverage, and at Galactic latitudes $|b| > 30\arcdeg$, 
to avoid large corrections for reddening.
Their parallax errors are $\sigma_{\pi}/\pi < 0.15$, and the parallax data are from the reduction by 
van Leeuwen (2007). The red clump stars are easily discernible at $M_K \sim -1.6$, and $ 0.5 \le (J-K)_0 \le 0.7$.
This color cut has been extensively used in studies of the thin and thick-disk parameters inferred from density laws 
(e.g., Cabrera-Lavers et al. 2005, 2007, Veltz et al. 2008), as well as more recent kinematical studies 
such as the estimation of the tilt of the velocity ellipsoid (Siebert et al. 2008). 

From Fig. 2, stars that contaminate this region are K dwarfs and subgiants.
Since K dwarfs are some 6 magnitudes fainter than red clump stars, they can be readily identified as objects with
very large velocities. K dwarfs have large proper motions as nearby stars, and wrongly adopted large distances, thus their velocities appear unusually large.
Of more concern however are the subgiants that reside in the same color range as red clump stars. 
To better understand their contribution in our sample, we inspect the surface gravities log~$g$ as a function of $(J-K)_0$ colors,
for the subsample that has stellar parameters determined. We show this plot in Figure 3. 
Red clump stars have surface gravities log~$g \sim 2$ to 3 (Puzeras et al. 2010 and references therein), and a color cut 
$0.5 \le (J-K)_0 \le 0.7$ clearly includes these stars, as well as a significant portion of subgiants and dwarf stars.
To minimize the contribution of subgiants, we thus restrict the color cut to $0.6 \le (J-K)_0 \le 0.7$, which 
helps exclude many of the stars with log~$g$ = 3 to 4. 

Next we inspect the total velocity in relation to log~$g$, to further eliminate dwarfs.
Velocities are determined from radial velocities, absolute proper motions and distances, where the distance
is derived from the absolute magnitude $M_K$ of red clump stars and the dereddened $K_0$ magnitude (see also next Section).
In Figure 4 we show the total velocity as a function of log~$g$ for the $0.5 \le (J-K)_0 \le 0.7$ sample (top), and
for the $0.6 \le (J-K)_0 \le 0.7$ (bottom). Objects with apparently large motions ($V_{tot} > 500$ km~s$^{-1}$) 
are clearly present at 
log~$g \ge 4.0$ representing nearby dwarfs with wrongly assigned distances. 
In the color range $0.5 \le (J-K)_0 \le 0.7$,
for log~$g = 2$ to 3, there appear to be 
a significant number of objects with relatively high velocities ($300 < V_{tot} < 500$ km s$^{-1}$),  
compared to those with log~$g \le 3$: these are most likely subgiants
contaminating the sample, that similarly to the dwarfs have been assigned too large distances.
In the color range $0.6 \le (J-K)_0 \le 0.7$, the population of likely subgiants appears much diminished.
For these reasons, in our subsequent analysis, we use red clump stars selected in the color range
$0.6 \le (J-K)_0 \le 0.7$. We further eliminate all stars with $V_{tot} \ge 400$ km s$^{-1}$, as likely nearby
stars with wrongly assigned distances.  With these cuts, the sample has 4815 objects, of which 
$45.2\%$ have log~$g$ determinations. We further impose one more restriction, i.e., for objects with
log~$g$ determinations, we keep only those that have log~$g \le 3.0$, amounting to a sample of 4420
stars to be analyzed (the fraction of objects with stellar parameters is now $40.3\%$). 
Next, we estimate the dwarf and subgiant contamination remaining in our entire sample. For the dwarfs this is done 
by determining the fraction of stars that have log~$g$ = [4,5] and $V_{tot} \ge 400$ km s$^{-1}$ in the
sample  with log~$g$ determinations. We proceed similarly for the subgiants, selecting this time stars 
with log~$g$ = (3,4).
These fractions are then scaled to the entire sample, and taking into account that $45.2\%$ of the stars were already
trimmed such that log~$g \le 3.0$. We obtain that the dwarf contribution to our sample of 4420 stars is $2\%$, 
while that of subgiants is $8\%$. These contamination estimates do not consider uncertainties in log~$g$.

In Figure 5 we show the $V$ and $K$ magnitude distributions of our entire sample (4420 stars), 
and the metallicity distribution, for
the subsample with stellar parameter determinations (1780 stars). The double peaked shape 
of the magnitude distribution is 
due to the input selection of RAVE stars (e.g., Fig 1 in Zwitter et al. 2008), and it will have bearing on
the contribution of thin disk stars. The metallicity distribution is indeed peaked toward high values,
ensuring that the absolute magnitude $M_K$ used for red clump stars is appropriate.
Our sample will therefore explore the kinematics of thin and thick disk red clump stars, and due  to the color 
selection will have no bearing on the
metal weak thick disk (e.g., Morrison et al. 1990) due to the color selection.

\section{Spatial Distribution and Velocities}

In Figure 6 we show the distribution of photometrically-determined distances from the Sun; 
the shape of the distribution is determined
by the input selection of RAVE stars (see also Fig. 5).
The spatial distribution of the sample is shown in Figure 7, where $(X, Y, Z)$ are Cartesian 
coordinates, with the Sun located at (8, 0, 0) kpc. $Y$ is positive toward Galactic rotation, and
$Z$ toward the north Galactic pole. We also define $R_{GC} = \sqrt{(X^2+Y^2)}$, as the distance from the GC, 
projected in the Galactic plane. Various panels show different projections. The higher concentration of
points near the Sun's location reflects once again the magnitude selection of the input list of RAVE stars, and
is not a spatial substructure. The bottom panels show the $XY$ distribution of the sample split into stars 
above and below the Galactic plane. Stars above the plane mainly occupy quadrant four, while stars below the 
plane occupy quadrant four and partly quadrants one and three. We note that the Galactic bar has its near end in
quadrant one, with the corotation radius between $R = 3.5 - 4.5 $ kpc 
and its orientation at $\sim 20\arcdeg$ from the GC-Sun direction (Gerhard 2010 and references therein).

Among the known inner-halo/thick-disk substructures that might affect our sample is the overdensity 
found in the first quadrant ($l = 20\arcdeg-40\arcdeg$) by Larsen \& Humphreys (1996) 
and further characterized by Parker et al. (2003, 2004),
Larsen et al. (2008, 2010). This overdensity is located between 1 to 2.5 kpc from the Sun, above and 
below the plane; however it does not extend into quadrant four. The galactic latitude extent
is from $|b| = 20\arcdeg-40\arcdeg$. Our sample avoids regions at $|b| < 25\arcdeg$ due to the RAVE input list
(see Fig. 1), and because we discard regions of high extinction.
Thus, our sample may be slightly affected by this overdensity below the plane in quadrant one. 

The Galactic stellar warp is known to have a starting Galactocentric radius of $\sim 8$ kpc 
(e.g. L\'{o}pez-Corredoira et al. 2002).
A recent model of the stellar warp by Reyl\'{e} et al. (2009) indicates
that the elevation of the disk midplane due to the warp is about 200 pc at
a galactocentric radius of 10 kpc, which is the largest Galactocentric radius encompassed by our sample.
At this radius however,
our sample includes stars more than 1 kpc from the plane (Fig. 7), so we do not believe 
that the kinematics of our sample is affected by the warp.

We calculate velocities in a cylindrical coordinate system ($V_R, V_{\theta}, V_z$), where
$V_R$ is positive outward from the GC, $V_{\theta}$ is positive in the Galactic rotation direction, and 
$V_z$ is positive toward the north Galactic pole. The velocities are in a Galactic rest frame, after
the rotation of the LSR, $V_{\theta}^0 = 220$ km~s$^{-1}$ and the solar 
peculiar motion ($V_R^\odot, V_{\theta}^\odot, V_z^\odot) = (-10.00,5.25,7.17)$ 
km~s$^{-1}$ (Dehnen \& Binney 1998) 
have been subtracted.
Velocities are derived from proper motions, radial velocities and distances. Errors in velocities are 
propagated from the errors in the proper motions, radial velocities and distances.
Proper-motion errors for the magnitude range explored here range between 0.4 and 4 mas~yr$^{-1}$, with an
average of 1.3 mas~yr$^{-1}$ (Girard et al. 2010). Radial-velocity errors
are between 0.3 and 4 km~s$^{-1}$, with
an average of 1.7 km~s$^{-1}$. Distance errors are determined from the error in 
the $K$ magnitude and the intrinsic scatter 
of 0.22 mags in the absolute magnitude of red clump stars. It is this latter number that dominates
the error budget, amounting to about $10\%$ in distance.
We have confirmed the propagated errors with errors derived 
via Monte Carlo simulations drawn from
errors in proper motions, radial velocities and distances.
The distribution of estimated velocity errors for each velocity component is shown in Figure 8.

\section{The LSR Velocity}

\subsection{Mean Velocities from This Study}

In Figure 9, we show the run of velocities as a function of distance from the Sun (top),
distance from the Galactic plane (middle), and distance from GC (bottom).
Left, middle and right rows show $V_R, V_\theta$ and $V_z$ respectively.
For $V_R$ and $V_z$ we also show a moving average to be compared with the expected value of 0 km~s$^{-1}$.
While $V_z$ is consistent with an average of 0 km~s$^{-1}$ for the entire $d_{sun}$ and
$|z|$ ranges, $V_R$ has a mean of 0 km~s$^{-1}$ only for stars within $\sim 1$ kpc from the Sun.
The slight departure of  $V_z$ from 0 km~s$^{-1}$ in the plot as a function of $R_{GC}$ at the 
extremes of the $R_{GC}$ range is due to small-number statistics at the edges, and thus is not significant.

For the entire sample, we obtain an average $V_R = 9.2 \pm 1.1$ km~s$^{-1}$, and
$V_z = -2.0 \pm 0.7$ km~s$^{-1}$. While $V_z$ is only marginally different from 0, $V_R$ is significantly
different from 0.
If we split the sample into a nearby subsample with $d_{sun} < 1.0$ kpc and a distant subsample
with $d_{sun} \ge 1.0$ kpc, we obtain: 1) $V_R^{near} = 3.0 \pm 1.7$  km~s$^{-1}$ and
$V_z^{near} = -2.6 \pm 1.0$ km~s$^{-1}$ for 1053 stars, and 2)
$V_R^{far} = 11.1 \pm 1.3$  km~s$^{-1}$ and
$V_z^{far} = -1.9 \pm 0.9$ km~s$^{-1}$ for 3367 stars.
Clearly the distant subsample behaves differently than the local one.
We have also split the entire sample of 4420 stars into subsamples above and below the plane to check
if a kinematical asymmetry is responsible for this non-zero $V_R$.
We obtain: $V_R^{above} = 14.1 \pm 1.9$  km~s$^{-1}$,
$V_z^{above} = -1.6 \pm 1.4$ km~s$^{-1}$ for 1225 stars above the plane, and $V_R^{below} = 7.3 \pm 1.3$, 
$V_z^{below} = -2.8 \pm 0.8$ km~s$^{-1}$ for 3195 stars below the plane.
The non-zero $V_R$ is significant in both samples. The below-the-plane sample, that may include 
in quadrant one part of the Larsen \& Humphreys (1996)  overdensity (see Section 3),  is thus not solely
responsible for the non-zero $V_R$.

To further explore the origin of this non-zero $V_R$, we have tested other values for the 
Solar peculiar motion and for LSR's rotation. Using the Schonrich et al. (2010) Solar peculiar 
motion, ($V_R^\odot, V_{\theta}^\odot, V_z^\odot) = (-11.10,12.24,7.25)$ km~s$^{-1}$, we obtain 
$V_R = 7.5 \pm 1.1$ km~s$^{-1}$, and $V_z = -2.0 \pm 0.7$ km~s$^{-1}$ for the entire sample.
Adopting the Schonrich et al. (2010) Solar peculiar motion, and the rotation of the LSR, 
$V_{\theta}^0 = 250$ km~s$^{-1}$
(Reid et al. 2009), we obtain $V_R = 5.0 \pm 1.1$ km~s$^{-1}$, and $V_z = -2.1 \pm 0.7$ km~s$^{-1}$.
Thus, $V_R$ is still at $\sim 5\sigma$ level different from 0.

Other possible sources for this unexpected result are of course systematics in the observed quantities, 
of which the most suspect are the proper motions and distances. First we test the distance, i.e., 
modify it in order to obtain an average $V_R = 0.0$ km~s$^{-1}$. We found that all distances have 
to be a factor of 0.5 smaller in order to
satisfy our requirement. This corresponds to a magnitude difference of 1.5, or the absolute magnitude of 
red clump stars has to be wrong by this amount, which seems unrealistic.
Next, we vary the proper motions in one coordinate e.g., RA (keeping the distance and proper motions in the 
other coordinate fixed), first by
adding an offset to it, and then introducing a slope with magnitude. 
The offset mimics an incorrect absolute proper motion correction, while the slope with magnitude 
mimics magnitude-dependent systematics
present in the proper motions. We found that we need an offset of 8 mas~yr$^{-1}$ in order to satisfy our 
requirement; in which case the averages are $V_R = 2.7 \pm 1.3$ km~s$^{-1}$, and $V_z = 4.6 \pm 0.8$ km~s$^{-1}$.
A slope of 3 mas~yr$^{-1}$~mag$^{-1}$ gives $V_R = 4.5 \pm 1.3$ km~s$^{-1}$, and $V_z = 4.1 \pm 0.9$ km~s$^{-1}$, neither being satisfactory. 
We repeat this test by varying the proper motion along Dec., and keeping the other quantities fixed.
Here, within the range explored, we find no satisfactory solution, because as $V_R$ nears 
low values, $V_z$ becomes $\sim 5\sigma$ different from zero.

We note that an offset of 8 mas~yr$^{-1}$ in the absolute zero point of SPM4 is a huge and unrealistic value,  since these
 systematics are expected to be of the order of 1-2 mas~yr$^{-1}$ (Girard et al. 2010).
Likewise a slope of 3 mas~yr$^{-1}$~mag$^{-1}$, is completely unacceptable for the SPM4 catalog, 
and in general for the SPM material and reductions. We remind the reader that our result 
for the absolute proper motion of 
cluster M 4 (Dinescu et al. 1999) obtained with SPM data is within 0.5 mas~yr$^{-1}$ of the result obtained 
with HST data (Bedin et al. 2003, Kalirai et al. 2004). The ground-based result uses data
in a much brighter magnitude range than the HST result: the first is tied to $Hipparcos$, the other to 
one QSO, and to galaxies. Thus, there is no reason to believe that systematics of $\sim 3$
 mas~yr$^{-1}$~mag$^{-1}$ are present in the SPM data.

To obtain acceptable values for $V_R$ {\bf and} $V_z$,
we need to have both proper-motion coordinates offset by 3  mas~yr$^{-1}$, or both with slopes of 
3 and 1 mas~yr$^{-1}$~mag$^{-1}$. Systematic errors this large in the SPM4 catalog are unrealistic.

\subsection{Other Results for $V_R$}

We turn now to an exploration of the literature regarding this issue. 
We consider only the most recent kinematical studies
that include large samples of stars, with well determined distances, proper motions and radial velocities,
and, when possible, homogeneous data.
One recent study that uses SDSS data in stripe 82 is that by Smith et al. (2009). That study 
uses a sample of 1700 halo subdwarfs, where radial velocities are from SDSS spectra, and the proper motions 
are determined only from SDSS data in stripe 82 which has repeated observations over a period of 7 years
(Bramich et al. 2008). Their derived values are $V_R = 8.9 \pm 2.6$ km~s$^{-1}$, and 
$V_z = -1.2 \pm 1.6$ km~s$^{-1}$,
for a sample spanning a heliocentric distance between 1 and 5 kpc.
This is in excellent agreement with our result. Smith et al. (2009) admit that this result for
$V_R$ is significantly different from the nominal zero, and suggest various causes for it such as
kinematical substructure; systematic errors in proper motions, distances or radial velocities; the presence
of binary stars. 

Another study is that of $\sim 1200$ metal-rich red giants at the South Galactic pole (SGP) 
done by Girard et al. (2006). This study 
used proper motions from an earlier version of the SPM catalog (namely SPM3, Girard et al. 2004), and 
photometric distances estimated from 2MASS photometry. At the SGP, the proper motions are directly projected
into $U,V$ velocities. While the study focuses on the shear of the thick disk, they obtained
an average velocity with respect to the Sun of $U = 19.1 \pm 2.7$ km~s$^{-1}$ at an average
2.2 kpc from the Galactic plane. Taking into account the Solar peculiar motion, their result is 
$U = 9.1 \pm 2.7$  km~s$^{-1}$, in agreement
with ours and with Smith et al. (2009). We note that the Girard et al. (2006) study used 
different tracers ($0.7 \le (J-K) \le 1.1$) than our current study.

Another study of red clump giants at the North Galactic Pole (NGP) was made by
Rybka \& Yatsenko (2009). The red clump stars were selected from 2MASS in the traditional color range
$0.5 \le (J-K) \le 0.7$, and the proper motions come from various catalogs including Tycho2 and UCAC2.
Using $\sim 1800$ stars
between 1 and 3 kpc from the Sun, they obtain an average velocity with respect to the LSR of
$U = 8.1 \pm 1.8$ km~s$^{-1}$, in agreement with our result, and the results mentioned above.

Two additional recent kinematical studies using SDSS DR7 data are those by 
C10 and B10. The C10 study focuses on ``calibration'' stars, or a subsample of the
SDSS data, while the B10 study uses all SDSS data. Both studies analyze dwarf stars (of the halo and disk), 
have the same photometric distance calibration described in Ivezi\'{c} et al. (2008), 
and use the same proper motions
determined from USNO-B and SDSS positions (Munn et al. 2004, 2008). Radial velocities are 
obtained from SDSS spectra (see details in C10 and B10).
Both studies have samples of tens of thousand of stars with heliocentric distances well outside 
the 1-kpc limit. Neither study gives the average $V_R, V_z$ for the entire sample, thus we can judge 
a small positive offset only from the plots presented.
Figure 6 in C10 shows histograms of each velocity component for various metallicity bins.
It is apparent that for $V_R$ the highest peak is always on the positive side of $V_R$ for all metallicity 
bins, while this is not the case for $V_z$. In metallicity bins -1.6 to -1.2 and -2.0 to -1.6,
this asymmetry is most apparent. 
However, for thick-disk stars selected in the metallicity range [Fe/H] = $(-0.8,  -0.6)$, and 
distance from the Galactic plane $|z| = (1, 2) $ kpc, C10 determine the average to be $V_R = 2.5 \pm 2.0$ km~s$^{-1}$.

B10 determine the average $V_R$ and $V_z$ for a 
subsample of ~13,000 M dwarfs located within $\sim 1$ kpc from the Sun, and obtain values consistent
with 0  km~s$^{-1}$. This is consistent with our results for objects within 1 kpc from the Sun.
However, in their Figure 7, where they show the run of the 
median $V_R$ as a function of distance from the plane, there appears to be a small positive offset
for stars between 1.5 and 3 kpc.

\subsection{Radial-velocity Studies Towards the Galactic Center and Anticenter}

Assuming the offset in $V_R$ is real, we believe that, rather than having the entire Galaxy expand,
it is more realistic to assume that stars in the solar neighborhood (within $\sim 1$ kpc from the Sun)
move inward, toward the GC compared to more distant stars that better
represent a Galactic rest frame.
If this is the case, then radial-velocity studies at low latitude and toward the Galactic center and 
anticenter should confirm/disprove this. 
In other words, bulge stars for instance should have an average radial velocity toward us, 
and stars at the anticenter (and beyond 1 kpc from the Sun) should have an average radial velocity away from us.
Izumiura et al. (1995) present a study of SiO masers toward the Galactic bulge, and they 
find an average shift in radial velocity of $-17.7 \pm 7.6$ km~s$^{-1}$. Taking into account the 
Solar peculiar motion, this leaves a net radial velocity of $\sim -7.7 \pm 7.6$ km~s$^{-1}$. 
Other studies that 
observe various types of stars toward the bulge, suffer from small sample size, 
large velocity dispersion of the bulge population, and
the modeling of its rotation.  One recent study that contains a significant number of  stars ($\sim 3300$) is
the radial-velocity study of M giants toward the bulge by 
Howard et al. (2008).
This study does not detect any net offset from zero for fields located along the minor axis
(their Fig. 14); however the scatter is rather large, of the order of 10 km~s$^{-1}$. Their subsequent 
study (Howard et al. 2009) of a stripe along the major axis at $b = -8\arcdeg$ 
with $\sim 1200$ red giants does show an offset of $-9.1 \pm 2.7$ km~s$^{-1}$ with respect to the 
LSR, which is intriguing, and in agreement with our result. It is unclear if this could be due 
to the asymmetry of the fields sampled along the major axis of the bulge combined with the 
rotation of the bulge.
Their radial-velocity histogram, however, has a Gaussian shape (their Fig. 2).

As for radial-velocity studies toward the Anticenter, we note that of Metzger \& Schechter (1994), who study 179
carbon stars at a distance of $\sim 6$ kpc from the Sun. They find that these stars move with a mean 
of $6.6\pm 1.7$ km~s$^{-1}$ with respect to the LSR, thus radially outward. Therefore this sample of stars
points yet again to the LSR's
motion toward the GC, with an amount similar to our result.

\subsection{Origin of the LSR Inward Motion}

Our investigation shows that local stars (lesser than $\sim 1$ kpc from the Sun)
do not exhibit a net $V_R$ motion, and thus 
$Hipparcos$-based results, for instance, are unlikely to detect this. 
However, more distant samples with full 3D velocities do detect it.

It is thus apparent that the entire local sample of stars
has a net inward motion toward the GC, while the vertical motion remains zero. 
The first thought that comes to mind is 
that this is due to some noncircular motion induced by perturbations
in the disk, such as spiral arms and/or bar. Indeed, a recent paper by Quillen et  al. (2010)
shows $UV$ maps of various neighborhoods that sample the disk of a N-body simulated galaxy.
These neighborhoods are placed at various radii from the galactic center, 
and various angles from 
the bar. The maps show velocity clumps and arcs induced by the bar and spiral pattern, that 
are offset from a mean of zero in velocity. Velocity offsets of these features can be 
easly as large as $\sim 50$ km s$^{-1}$. Thus our offset of $\sim 12$ km s$^{-1}$ for the
entire local solar neighborhood is quite within the ranges predicted in Quillen et  al. (2010).

\section{Velocity and Velocity-Dispersion Gradients}

\subsection{Variation with Distance from the Galactic Plane}

In Figure 10 we show the run of each velocity component as a function of
distance from the Galactic plane $|z|$; the averages in the top panel and the
dispersions in the bottom panel. The data are grouped in 
bins of equal number of stars, here with 124 stars per bin. 
The data are restricted to a Galactocentric 
radius range of $ 7.5 < R_{GC} < 8.5$ kpc. 
The average in each bin is calculated using 
probability plots (Hamaker 1978) trimmed at $10\%$ on each side, and the 
uncertainty in the average is derived from the dispersion, also determined from probability plots.
The bottom plot of Fig. 10 shows the {\it intrinsic} velocity dispersions, i.e., corrected for
velocity errors.
Intrinsic velocity dispersions are calculated as follows. 
We start with an initial guess of the intrinsic dispersion. For each point we calculate
the square of the quadrature sum of this intrinsic dispersion and
its velocity error, and we divide this number with the point's deviation from the mean.
For all points this ratio should have a Gaussian distribution with 
a standard deviation of 1, if the intrinsic dispersion is correct. If not, the initial intrinsic 
dispersion is adjusted and the procedure is repeated until we satisfy the above condition.

The top plot shows the variation of the rotation velocity $V_{\theta}$ as a function of $|z|$;
of $V_R$, which is offset from 0 km~s$^{-1}$ at z-distances larger than $\sim 0.7$ kpc as
discussed in the previous Section; and of $V_z$, which is consistent with 0 km~s$^{-1}$.

The bottom plot, which displays the velocity dispersions as a function of $|z|$, 
clearly shows the transition between the 
thin and thick disks. At $|z| = 0.5 $ kpc, the sample is dominated by thin-disk red clump stars reflected
in the low dispersion in all three components: $(\sigma_{V_R},\sigma_{V_{\theta}},\sigma_{V_z}) =
(45, 30, 25)$ km~s$^{-1}$. This dispersion increases rapidly between $|z| = 0.5$ and 1 kpc
in all three components, to reach values of $(\sigma_{V_R},\sigma_{V_{\theta}},\sigma_{V_z}) \sim
(70, 48, 38)$ km~s$^{-1}$ at 1 kpc from the plane. This rapid increase reflects the 
mixture between the thin disk population with its low velocity dispersion and the  thick disk
population with a large dispersion. Before correcting for the contribution of the thin disk,
we would like to discuss the results of three other studies.
The continuous lines with corresponding colors for each velocity component show the 
dispersion dependence on $|z|$ as determined by B10 for SDSS data for disk, dwarf stars.
These stars were selected to be among the ``blue stars'', specifically
with $0.2 < (g-r) < 0.6$, and $0.7 < (u-g) < 2.0$ (and two other criteria for combined 
colors, see B10), as well as to have photometrically-determined metallicities 
[Fe/H]~$ > -0.9$. Thus the metal rich end
is determined by the color cuts, and the sample is thought to represent the thick disk according
to metallicity. In all three velocity components, the dispersions determined by B10, are 
substantially smaller than our values, except at $|z| \sim 0.5$ kpc. In fact, the B10 values 
appear to reflect the kinematics of the thin disk rather than the thick disk.
In the same plot, we show the run of velocity dispersions determined by Girard et al. (2006)
at the SGP from the analysis of metal-rich, thick-disk giants. These values are more in line 
with what we obtain at $|z| > 1$ kpc; however our values have yet to be corrected for the
contamination of the thin disk contribution, a correction that will push the dispersion 
to higher values.

We add one more estimate of the dispersions, that determined by C10 from SDSS data for dwarf stars.
Their thick-disk sample is selected to be in the metallicity range [Fe/H] = (-0.8, -0.6),
where metallicities are determined from SDSS spectra.
The dispersions are estimated for stars within $|z| = 1 - 2$ kpc. At an average $|z| = 1.1$ kpc,
they obtain $(\sigma_{V_R},\sigma_{V_{\theta}},\sigma_{V_z}) =
(53, 51, 35)\pm(2, 1, 1) $ km~s$^{-1}$. These values are not corrected for 
observational errors in the velocities, which are assumed to be on the order of $5-6$ km~s$^{-1}$ (C10).
We represent the C10 dispersions with a triangle symbol in our Fig. 9.
While their dispersions in the rotation and vertical component agree with ours, and are larger than
the B10 values, the dispersion in the radial direction is substantially smaller than our value.
It is also somewhat intriguing that the radial velocity dispersion is practically equal to
the rotation velocity dispersion, clearly different from other thick-disk studies where 
the radial dispersion is larger by 20 to $30\%$.

In light of the results summarized here, the B10 velocity dispersions seem unusually low, and it is
worth exploring this a bit further.
The C10 study used color cuts of ``bluer'' limits than B10, specifically,
$0.0 < (g-r) < 0.6$, and $0.6 < (u-g) < 1.2$, and a metallicity range with a cutoff 
at the high end. Their dispersions are larger than those in B10. It is thus likely that the
B10 sample of ``disk'' stars is significantly contaminated by thin disk stars.

Next we determine the velocity and velocity dispersion gradients, taking into account 
the contamination by thin-disk stars.
As a first guess for the contribution of the thin disk in our sample,
we use the density laws determined by Cabrera-Lavers et al. (2005, hereafter C05) from
starcounts of red clump stars in 2MASS. The density laws are expressed as exponential
functions of $|z|$, with given scale heights for the thin and thick disks, and a given
normalization of thick to thin disk stars.
We show two parametrizations: 1) $z_{thin} = 269$ pc, $z_{thick} = 1062$ pc, and
a local normalization of $8.6\%$ thick disk stars to thin disk, and 2)
$z_{thin} = 225$ pc, $z_{thick} = 1065$ pc, and
a local normalization of $11.4\%$ thick disk stars to thin disk. In Figure 11 we show
the fraction of thin disk stars (in $\%$) as a function of $|z|$ for the two descriptions.
At $\sim 1$ kpc, these descriptions indicate that the contribution of thin disk is between
$20\%$ and $40\%$ of the total stars. Our sample has a magnitude-selection imposed by
the input RAVE catalog that affects this ratio that was derived for complete samples.
Instead we choose to derive this ratio from our data by modeling the distribution of 
$V_{\theta}$ as the sum of two Gaussians: one representing the thin disk, the other the thick disk.
The mean velocity of the thin disk is fixed at 220 km~s$^{-1}$. 
We determine via $\chi^2$ minimization of the $V_{\theta}$ distribution the velocity dispersions of the 
thin and thick disks, the fraction of thin disk stars, and the mean velocity of the thick disk.
We divide our sample into three $|z|$ bins: a) from 0.3 to 0.5 kpc, b) from 0.5 to 1.0 kpc, and
c) from 1.0 to 1.5 kpc and apply this procedure for each bin.
In Table 1 we show the best-fit results for each bin.

\begin{table}[htb]
\caption{Thin-disk Fraction from $V_{\theta}$ Distribution}
\begin{tabular}{rrrrrr}
\tableline
\\
\multicolumn{1}{c}{$\overline{z}$} &
\multicolumn{1}{c}{$N_{all}$} &
\multicolumn{1}{c}{$f_{thin}$} &
\multicolumn{1}{c}{$\sigma_{\theta}^{thin}$} &
\multicolumn{1}{c}{$\overline{V_{\theta}^{thick}}$} &
\multicolumn{1}{c}{$\sigma_{\theta}^{thick}$} \\
\multicolumn{1}{c}{(kpc)} &
\multicolumn{1}{c}{} &
\multicolumn{1}{c}{($\%$)} &
\multicolumn{1}{c}{(km~s$^{-1}$)} &
\multicolumn{1}{c}{(km~s$^{-1}$)} &
\multicolumn{1}{c}{(km~s$^{-1}$)} \\
\tableline
\\
\multicolumn{1}{r}{0.4} & \multicolumn{1}{r}{426} &  \multicolumn{1}{r}{$35\pm42$} & \multicolumn{1}{r}{$19\pm5$} & \multicolumn{1}{r}{$194\pm17$} & \multicolumn{1}{r}{$29\pm5$} \\
\multicolumn{1}{r}{0.7} & \multicolumn{1}{r}{805} &  \multicolumn{1}{r}{$24\pm6$} & \multicolumn{1}{r}{$18\pm2$} & \multicolumn{1}{r}{$185\pm03$} & \multicolumn{1}{r}{$35\pm1$} \\
\multicolumn{1}{r}{1.2} & \multicolumn{1}{r}{359} &  \multicolumn{1}{r}{$12\pm2$} & \multicolumn{1}{r}{$16\pm2$} & \multicolumn{1}{r}{$175\pm02$} & \multicolumn{1}{r}{$50\pm1$} \\
\tableline
\end{tabular}
\end{table}

The first z-bin is poorly constrained, due to the proximity of the rotation velocity of the thick disk to that of the thin disk.
The next two z-bins have reasonable fits, with the middle one probably the best, due to the number of stars included.
If we use the two C05 laws, but normalize such that the thin-to-thick disk ratio is set 
by the $\overline{z} = 0.7$ kpc result from Tab. 1, we obtain the dashed lines shown in Fig. 11.
This indicates that the thin-disk contamination
is lower than that based directly on the  C05 laws.
Determinations from all three bins shown in Tab. 1, are also plotted in Fig. 11. 
The most distant point from the plane 
is in reasonable agreement with 
the predicted values, while the nearest is simply a poor constraint. 
Since the two C05 laws are very similar
with this new normalization at $|z| > \sim 0.5$ kpc, we will adopt only one to correct the velocity dispersions,
namely the one with the higher scale height (dashed black line in Fig. 11).

We correct the measured dispersions due to the contamination of the thin disk in the following way.
The measured dispersion is:

\begin{equation}
\sigma_m^2 = a\times\sigma_a^2+b\times\sigma_b^2
\end{equation}
where, $a$ is the fraction of thin disk stars, and $\sigma_a$ is the dispersion of 
thin disk stars, while $b$ is the fraction of thick disk stars, and $\sigma_b$ is the 
dispersion of thick disk stars.
We derive $\sigma_b$ by adopting the fraction of thin and thick disk stars at each $|z|$
from the relationship presented above, and adopting the following dispersions for the old,
thin disk: $(\sigma_{V_R},\sigma_{V_{\theta}},\sigma_{V_z}) =
(40, 24, 20)$ km~s$^{-1}$ (Nordstrom et al. 2004).
This correction is applicable if the mean velocities for thin and thick disk are the same,
and thus can be applied in principle only to the $V_R$ and $V_z$ components.
To understand the influence of thin-disk contamination on $V_{\theta}$, we use Monte Carlo simulations
to generate a population of thin and thick disk stars. 

A set of 10000 points are generated
at a series of $|z|$ values.
Velocities are drawn from Gaussian distributions
of given means and dispersions that resemble the thin and thick disks.  The thick disk rotation velocity
varies linearly with $|z|$, with a gradient similar to that seen in observations.
Velocities obtained this way are then
smoothed with velocity errors, these being drawn from the observed distribution of errors in $V_{\theta}$.
The fraction of thin disk stars as a function of $|z|$ is that given by the C05 law, normalized with our data.
We then calculate the average and the dispersion at each $|z|$ value, applying the same procedure as 
with the observed data, i.e., probability plots for averages, and the intrinsic dispersion routine that eliminates
measurement errors from the observed dispersions. We then compare these determined values with those 
from the input. We find that the dispersion correction is small (less than 2  km~s$^{-1}$) for
$|z| > 0.7$ kpc, and similar to the correction given by equation 1. This is probably due to the
fact that the thin disk fraction is small at this distance from the plane ($\sim 20\%$ at $|z| \sim 0.8$ kpc, 
see Fig. 11). However, the average velocity is affected by the thin-disk contamination, well beyond
$|z| \sim 0.7$.

With this procedure, we determine the gradients of the velocity dispersions of the thick disk by applying eq. 1  
and using only points with $|z| > 0.7$ kpc, i.e., only seven bins,  which are fit with a line. The fits are shown
in Figure 12 (black lines) for each velocity component, together with our original data (grey symbols) and those
corrected for thin-disk contamination (black symbols). The linear dependence determined by Girard et al. (2006) is 
shown with a red line; the thin-disk fraction is represented with a blue line, and the dashed line
in the top plot represents one of the equilibrium models using the Jeans equation calculated 
in Girard et al. (2006) that best fit their data (namely model 4, their Table 1 and Figure 8). The vertical dotted line
indicates the limit $|z| = 0.7$ kpc.
The results are summarized in Table 2.

\begin{table}[htb]
\caption{Velocity Dispersion Gradients with $|z|$}
\begin{tabular}{rrr}
\tableline
\\
 & \multicolumn{1}{c}{$\sigma\vert_{z=1.0}$} &
  \multicolumn{1}{c}{$\partial \sigma / \partial z$} \\
 & \multicolumn{1}{c}{(km~s$^{-1}$)} &
  \multicolumn{1}{c}{(km~s$^{-1}$~kpc$^{-1}$)} \\
\tableline
\multicolumn{1}{c}{$V_R:~$} & \multicolumn{1}{c}{$70.4\pm4.1$} & \multicolumn{1}{c}{$17.4\pm2.5$} \\
\multicolumn{1}{c}{$V_{\theta}:~$} & \multicolumn{1}{c}{$48.0\pm8.3$} & \multicolumn{1}{c}{$17.1\pm5.0$} \\
\multicolumn{1}{c}{$V_z:~$} & \multicolumn{1}{c}{$36.2\pm4.0$} & \multicolumn{1}{c}{$5.0\pm2.4$} \\
\tableline
\end{tabular}
\end{table}

For both $V_R$ and $V_z$ our dispersion gradients are larger than those determined by Girard et al. (2006),
which are $7.5\pm3.1$ km~s$^{-1}$ and $10.5\pm3.3$ km~s$^{-1}$ respectively. This however does not represent disagreement,
since the actual dependence is not linear. Girard et al. (2006) calculate several theoretical profiles
which show that these dependencies are better approximated with quadratic functions. However they can be approximated
with linear functions over limited $|z|$ ranges. Toward low $|z|$ values, the profiles become steeper: thus 
for the range in Girard et al. (2006) from 1 to 4 kpc, the dispersion gradient is shallower than ours  where the 
range is from 0.7 to 2.5 kpc.

Finally, we estimate the gradient of $V_{\theta}$ as a function of $|z|$.
Our simulations show that the average $V_{\theta}$ is affected by the thin-disk contamination 
out to $|z| \sim 1.1$ kpc, where the bias is about the size of the error in the average velocity
(i.e. $\sim 5$   km~s$^{-1}$). Only three bins in our dataset are beyond this $|z|$ value.
Thus, based on our simulations, we derive the bias in the average velocity at each $|z|$ bin.
We apply this bias to all of our our data bins, and then fit a line to obtain a gradient 
$\partial V_{\theta} / \partial z = -25.2 \pm 2.1$ km~s$^{-1}$, with $V_{\theta}\vert_{z=0} = 200.8\pm2.1$ km~s$^{-1}$.
Since we are uncertain of the thin-disk fraction at low 
$|z|$ values, and therefore of the correction to be applied, we provide yet another estimate of the gradient. 
We use only the outermost three data points
from the binned data, and two points obtained from the two-Gaussian fit of $V_{\theta}$ distribution, those at
$|z| = 0.7$ and 1.2 kpc (see Tab. 1). The gradient obtained this way is $\partial V_{\theta} / \partial z = -24.4 \pm 1.6$ km~s$^{-1}$,
and $V_{\theta}\vert_{z=0} = 202.3\pm2.4$ km~s$^{-1}$, and in agreement with the previous determination.
We also note that the gradient obtained using all bins, and no bias correction for thin-disk contamination is
$\partial V_R / \partial z = -30.9 \pm 2.1$ km~s$^{-1}$, indicating that the contamination tends to steepen the gradient.
Our determination is less steep than that determined by Girard et al. (2006) of $-30.3 \pm 3.2$ km~s$^{-1}$; however it is
within $1\sigma$ error. The C10 determination of $-36 \pm 1$ km~s$^{-1}$ is significantly different from ours.

We also calculate $V_R$ and $V_z$ as given by the average over bins at $|z| > 0.7$ kpc, to avoid the issue associated with a local sample of
stars as discussed in Section 4. These values are $\overline V_R = 12.7\pm3.1$ km~s$^{-1}$, and $\overline V_z = -0.7\pm1.9$ km~s$^{-1}$.

\subsection{Variation with Distance from the Galactic Center}

To study the dependence with $R_{GC}$, we first trim our sample to include stars
at $|z| > 1.0$ kpc. By doing so we avoid the contamination of the thin disk, while  still having a representative 
number of stars. We then define three layers with $1.0 < |z| < 1.5$ kpc (1115 stars) , $1.5 < |z| < 2.0$ kpc (401 stars), 
and $2.0 < |z| < 2.5$ kpc (190 stars).
In each of these layers, we calculate the velocity averages and intrinsic dispersions 
in four equal-number $R_{GC}$ bins  
using the same procedures as in Section 5.1. 
The velocity average and dispersions as a function of $R_{GC}$ are shown in Figure 13, for each velocity component.
Each $z$-layer is represented with a different symbol: filled circles for $1.0 < |z| < 1.5$ kpc, open circles for
 $1.5 < |z| < 2.0$ kpc,
and open triangles for $2.0 < |z| < 2.5$ kpc. Here, velocity dispersions are not corrected for thin-disk 
contamination, which is very small and essentially similar in each $z$-layer.

Inspection of Fig. 13 shows that velocity averages do not show significant gradients with $R_{GC}$, except 
$V_{\theta}$ which decreases mildly with decreasing distance form GC. 
As for dispersions, the only significant gradient with $R_{GC}$ is that for
$\sigma_{V_z}$. 
In Table 3 we list the gradients for each velocity component, and each $z$-layer. These were obtained
using linear fits to the data. Note that the middle layer spans the largest $R_{GC}$ range.

\begin{table}[htb]
\caption{Velocity and Velocity Dispersion Gradients with $R_{GC}$}
\begin{tabular}{rrrrrrrrr}
\tableline
\\
\multicolumn{1}{c}{$z$-range} & \multicolumn{1}{c}{N} & \multicolumn{1}{c}{$\partial V_R/\partial R_{GC}$} & \multicolumn{1}{c}{$\partial V_{\theta}/\partial R_{GC}$} &\multicolumn{1}{c}{$\partial V_z/\partial R_{GC}$} & & \multicolumn{1}{c}{$\partial \sigma_{V_R} / \partial R_{GC}$}  &
 \multicolumn{1}{c}{$\partial \sigma_{V_{\theta}} / \partial R_{GC}$}  &
 \multicolumn{1}{c}{$\partial \sigma_{V_z} / \partial R_{GC}$}  \\
 \multicolumn{1}{c}{(kpc)} & & \multicolumn{3}{c}{(km~s$^{-1}$~kpc$^{-1}$)} & & \multicolumn{3}{c}{(km~s$^{-1}$~kpc$^{-1}$)} \\
\tableline
\multicolumn{1}{c}{1.0-1.5} & \multicolumn{1}{c}{1112} & $-1.0\pm 2.1$ & $6.0\pm1.4$ & $1.3\pm1.8$ & & \multicolumn{1}{c}{$-1.6\pm3.2$} &  \multicolumn{1}{c}{$-1.7\pm4.2$} &  \multicolumn{1}{c}{$-7.5\pm1.3$} \\
\multicolumn{1}{c}{1.5-2.0} & \multicolumn{1}{c}{400} & $-0.5\pm 5.0$ & $13.0\pm4.5$ & $6.6\pm4.6$ & & \multicolumn{1}{c}{$-1.6\pm6.3$} &  \multicolumn{1}{c}{$-1.0\pm4.3$} &  \multicolumn{1}{c}{$-8.3\pm4.7$} \\
\multicolumn{1}{c}{2.0-2.5} & \multicolumn{1}{c}{188} & $-2.5\pm 4.5$ & $12.0\pm8.1$ & $2.1\pm1.2$  & & \multicolumn{1}{c}{$-3.4\pm2.9$} &  \multicolumn{1}{c}{$-6.0\pm3.5$} &  \multicolumn{1}{c}{$-8.4\pm3.1$} \\
\tableline
\end{tabular}
\end{table}

The gradient $\partial \sigma_{V_z} / \partial R_{GC}$ is related to the underlying gravitational potential, and provides
a dynamical estimate of the thin disk scale length to be discussed in Section 7.

\section{Tilt Angles}

The tilt of the velocity ellipsoid in cylindrical coordinates is described by the correlation coefficient:
\begin{equation}
C[V_R,V_z] = \frac{\sigma^2_{V_{R,z}}}{(\sigma^2_{V_R} \sigma^2_{V_z})^{1/2}}
\end{equation}
and by the tilt angle:
\begin{equation}
\tan (2\alpha_{Rz}) = \frac{2\sigma^2_{V_{R,z}}}{(\sigma^2_{V_R}-\sigma^2_{V_z})}
\end{equation}
where
\begin{equation}
\sigma^2_{V_{R,z}} = \frac{1}{(N-1)}\displaystyle\sum\limits_{i=1}^N (V_{R,i}-\overline{V_R}) (V_{z,i}-\overline{V_z})
\end{equation}
and
\begin{equation}
\sigma^2_{V_R} = \frac{1}{(N-1)}\displaystyle\sum\limits_{i=1}^N (V_{R,i}-\overline{V_R})^2;
~~~~~\sigma^2_{V_z} = \frac{1}{(N-1)}\displaystyle\sum\limits_{i=1}^N (V_{z,i}-\overline{V_z})^2
\end{equation}

In these equations, the pair $(R,z)$ can be replaced with $(R,\theta)$ and $(\theta,z)$, for the other two components.
The cross term and dispersions (eqs. 4 and 5) are calculated here from the data, without any attempt to 
subtract measurement errors in velocities. Instead the contribution of these errors will be modeled via Monte Carlo
simulations. Figure 8 shows that the distributions of velocity errors vary by component. This will affect the estimation of the tilt angles. For example, 
the $V_R$ errors are on average larger than the $V_z$ errors, a fact that will bias the $\alpha_{Rz}$ angle 
to smaller values (see also Siebert et al. 2008).

First, we choose a sample of stars with  $7.0 < R_{GC} < 9.0$ kpc, i.e., located within a cylindrical shell 
that includes the Sun. We determine the tilt angle in two samples: above the plane with $ 0.7 < z < 2$ kpc including 387 stars,
and below the plane with $ -2 < z < -0.7$ kpc including 1063 stars. 
The velocities are shown in Figure 14, with the above-the plane sample shown in the left panels, and the below-the-plane sample in the
right panels. 
We determine the tilt angle (eq. 3), the correlation coefficient (eq. 2), and the ratio of the
minor and major axes of the velocity ellipsoid for both samples and all velocity pairs. These are listed in Table 4.
For the below-the-plane sample,  
the expected symmetry with respect to the Galactic plane is taken into account in the values of the angle and correlation coefficient
shown in Tab. 4. That is, for the below-the-plane sample, the sign of the $V_{z}$ component is flipped. 

Uncertainties in these values are calculated as follows.
First, we determine the contribution of measurement errors by generating datasets with 
proper motions, radial velocities and distances drawn from the Gaussian distribution of their errors. The scatter obtained 
in the velocity ellipsoid parameters from these datasets represent the contribution of velocity measurement error. 
Second, we determine the uncertainties in the velocity ellipsoid parameters due to the finite number of data points used in the 
determination. Since the intrinsic velocity dispersions are considerably larger than the velocity measurement errors, 
this second contribution is substantial. We generate velocity ellipsoids with
given axis ratios, and calculate the ellipsoid's parameters using the same number of points as in 
our observed samples; the scatter in parameters is adopted as the second contribution to the errors.
 These two different contributions are added in quadrature to obtain the final
uncertainties in the ellipsoid parameters.

\begin{table}[tbh]
\caption{Tilt angles, correlation coefficients and axis ratios}
\begin{tabular}{rrrrrrrr}
\tableline
\\
 & \multicolumn{3}{c}{above the plane ($N=387$)} & & \multicolumn{3}{c}{below the plane ($N=1063$) } \\
\tableline
 & \multicolumn{1}{c}{$\alpha$} & \multicolumn{1}{c}{C} & \multicolumn{1}{c}{ratio} &  &\multicolumn{1}{c}{$\alpha$} & \multicolumn{1}{c}{C} & \multicolumn{1}{c}{ratio} \\
& \multicolumn{1}{c}{($\arcdeg$)} & & & & \multicolumn{1}{c}{($\arcdeg$)} & & \\
\tableline
\multicolumn{1}{c}{$Rz:$} & \multicolumn{1}{r}{$6.3\pm4.9$} & \multicolumn{1}{r}{$0.077\pm0.060$} & \multicolumn{1}{c}{$0.71\pm0.05$} & & \multicolumn{1}{r}{$8.8\pm1.9$} & \multicolumn{1}{r}{$0.163\pm0.035$} & \multicolumn{1}{c}{$0.59\pm0.03$} \\ 
\multicolumn{1}{c}{$R \theta:$} & \multicolumn{1}{r}{$-10.3\pm4.3$} & \multicolumn{1}{r}{$-0.134\pm0.056$} & \multicolumn{1}{c}{$0.69\pm0.04$} & & \multicolumn{1}{r}{$-6.7\pm4.6$} & \multicolumn{1}{r}{$-0.056\pm0.037$} & \multicolumn{1}{c}{$0.79\pm0.04$} \\ 
\multicolumn{1}{c}{$\theta z:$} & \multicolumn{1}{r}{$48.8\pm13.8$} & \multicolumn{1}{r}{$0.123\pm0.035$} & \multicolumn{1}{c}{$0.88\pm0.05$} & & \multicolumn{1}{r}{$5.7\pm3.7$} & \multicolumn{1}{r}{$0.054\pm0.035$} & \multicolumn{1}{c}{$0.76\pm0.03$} \\ 
\tableline
\end{tabular}
\end{table}

From Table 4, the tilt angle with 
the most significant nonzero value is $\alpha_{Rz}$ in the sample below the plane.
The next most significant is $\alpha_{\theta z}$ in the sample above the plane.
The axis ratio in general indicates
how flattened each ellipsoid is; for a nearly-circular distribution (ratio $\sim 1$), 
the tilt angle can vary substantially indicating that 
it is practically undetermined. 

The most circular projection is $\theta z$ in the above-the-plane sample, explaining why 
this angle is the most ill-constrained.
Inspection of Fig. 14 (bottom-left), reveals a rather unusual velocity distribution 
for this component/sample when compared to the others: it is asymmetric and not well 
described by an ellipse and its core appears less tilted than the outer 
regions. This unusual distribution may be due to a real kinematical structure
of accreted or resonant origin; however, the data available are insufficient to reliably
test this. 
We have therefore chosen to disregard this particular angle estimate.
Our final angles are therefore determined from the combined above- and below-the-plane samples, 
except for the 
pair $\theta z$, where the determination is only from the below-the-plane sample. The combined sample has 1450 stars with 
$(\overline{R_{GC}},\overline{z}) = (7.8, 1.1)$ kpc.

Next, we determine the bias in the tilt angle introduced by velocity errors. 
We run Monte Carlo simulations adopting a thick disk with intrinsic velocity dispersions of 70 and 40 km~s$^{-1}$,
along a major and minor axis respectively, and a range of input tilt angles from $0\arcdeg$ to $25\arcdeg$.
Velocities projected onto $V_R$ and $V_z$ are then convolved with velocity errors drawn from the observed 
error distributions. Tilt angles are then calculated according to eqs. 2-5, and compared to the input values.
As in Siebert et al. (2008), we find that the measured angle is slightly smaller than the true one. However, our
bias is not as large as that in Siebert et al. (2008) because our intrinsic dispersions are large
compared to the velocity errors. In Siebert et al. (2008), intrinsic velocity dispersions are smaller, 
as they sample red clump stars within $ 0.5 < |z| < 1.5$ kpc, where the thin disk is contributing significantly.
Also, by looking only at the SGP, their velocity-error distribution is quite different in the $z$ direction from the other two,
because radial velocity errors contribute only to $V_z$ while proper-motion errors contribute to $V_{R,\theta}$.
Our relationships for the intrinsic tilt angle and correlation coefficient are:
\begin{equation}
\Delta \alpha_{true-meas} = 0.073(\pm0.009) \times  \alpha_{meas};~~~
\Delta C_{true-meas} = 0.078(\pm0.014) \times C_{meas}
\end{equation}

In Table 5 we list the derived tilt angles, correlation coefficients and axis ratios. 
Only the $R,z$ quantities were corrected for bias, since the other two are insensitive to it.

\begin{table}[tbh]
\caption{Tilt angles and correlation coefficients}
\begin{tabular}{rrrr}
\tableline
\\
& \multicolumn{1}{c}{$\alpha$} & \multicolumn{1}{c}{$C$} & \multicolumn{1}{c}{ratio}\\
& \multicolumn{1}{c}{($\arcdeg$)} &  & \\
\tableline
$R z$: & $8.6\pm1.8$ & $0.141\pm0.029$ & $0.62\pm0.03$ \\
$R \theta$: & $-8.2\pm3.2$ & $-0.079\pm0.031$ &  $0.76\pm0.03$ \\
$\theta z$: & $5.7\pm3.7$ & $0.054\pm0.035$  & $0.76\pm0.03 $ \\
\tableline
\end{tabular}
\end{table}

The angle $\alpha_{Rz}$ we have derived is in agreement with the C10 determination 
for metal-rich stars within $ 1 < |z| < 2$, of $7.1\arcdeg \pm 1.5\arcdeg$, and
with that determined by Siebert et al. (2008) of $7.3\arcdeg \pm 1.8\arcdeg$.

\section{Scale Length of the Thin Disk}

\subsection{Formulation}

We derive the profile of $\sigma_{V_z}$ as a function of $R_{GC}$ for a relaxed 
thick-disk population in equilibrium within the underlying static gravitational potential of the Galaxy.
We use the Jeans equation:

\begin{equation}
\frac{\partial(\rho \overline{V_R V_z})}{\partial R} + \frac{\partial (\rho\overline{V_z^2})}{\partial z} 
+ \frac{\rho \overline{V_R V_z}}{R} = - \rho \frac{\partial \Phi_{tot}}{\partial z}
\end{equation}

Here, $R$ and $z$ are Galactocentric cylindrical coordinates, $\rho(R, z)$ is the 
volume density of the thin-disk stars, and $\Phi_{tot}(R,z) $ is the gravitational potential.
Also, the velocity dispersions and cross term are:

\begin{equation}
\overline{V_R^2} = \sigma^2_{V_R},~~~~
\overline{V_z^2} = \sigma^2_{V_z},~~~~
\overline{V_R V_z} = \sigma^2_{V_{Rz}} \\
\end{equation} 

We adopt for the cross term the
expression derived by Binney \& Tremaine (1987, page 199) in terms of the limiting case of spherical alignment:
\begin{equation}
\overline{V_R V_z} = (\overline{V_R^2} - \overline{V_z^2}) \frac{z}{R} = \overline{V_z^2}~\beta~\frac{z}{R}
\end{equation}
$\beta$ is a dimensionless number between 0 and 1, with 0 for cylindrical alignment of the velocity ellipsoid, 
and 1 for spherical alignment. The density distribution of the thick disk population is represented by 
exponential functions in radial and vertical directions:
\begin{equation}
\rho(R,z) = \rho_0~ \exp~ \Bigg(- \frac{R}{R_{thick}} - \frac{z}{z_{thick}} \Bigg)
\end{equation}
where $R_{thick}$ and $z_{thick}$ are the scale length and scale height of the thick disk.
Using the formulations of  eq. (8) and (9) in eq. (6),  we obtain:
\begin{equation}
\overline{V_z^2(R,z)} = \frac{1}{\frac{z~\beta}{R~R_{thick}} + \frac{1}{z_{thick}}}~ \Bigg( \frac{\partial \Phi_{tot}}{\partial z} - \frac{z~\beta}{R}~\frac{\partial \overline{V_z^2}}{\partial R} \Bigg)
\end{equation}
The Galactic potential separated by component is:
\begin{equation}
\Phi_{tot}= \Phi_{bulge} + \Phi_{disk} + \Phi_{halo} 
\end{equation}
The bulge potential adopted from the formulation of Johnston et al. (1995) is:
\begin{equation}
\Phi_{bulge} = - \frac{G M_b}{r+c}
\end{equation}
where, $r = \sqrt{R^2+z^2}$, $c=0.7$ kpc, $G$ is the gravitational constant,
and $M_b = 3.4 \times 10^{10} M_{\odot}$ is the mass of the bulge.
The disk potential, representing here the thin disk, is expressed with Bessel functions as in Girard et al. (2006). 
This potential, with an exponential surface density of the form $\Sigma_{thin}(R) = \Sigma_0 \exp(-R/R_{thin})$, is:
\begin{equation}
\Phi_{disk} = -2\pi G \Sigma_0 R_{thin}^2 \times \int_0^{\infty} \frac{J_0(kR) \exp(-k|z|) dk}{(1+k^2R_{thin}^2)^{3/2}}
\end{equation}
where $R_{thin}$ is the scale length of the thin disk.
The halo potential is a Plummer model given by:
\begin{equation}
\Phi_{halo} = - \frac{G~M_h}{(r^2+a^2)^{1/2}}
\end{equation}
where $a = 6.3$ kpc is the core radius, and $M_h$ is the halo mass.
The halo mass is expressed in terms of the rotational velocity of the LSR, $V_c$ as:
\begin{equation}
V_c^2 = \frac{G M_h R_{\odot}^2}{(R_{\odot}^2+a^2)^{3/2}} + V_{disk}^2(R_{\odot},0) + V_{bulge}^2(R_{\odot},0)
\end{equation}
where $V_{disk}$ and $V_{bulge}$ represent the rotation associated with the thin-disk gravitational potential
and the bulge potential respectively. These can be expressed as:
\begin{equation}
V_{disk}^2(R,z) = R\frac{\partial \Phi_{disk}}{\partial R} = 2 \pi G \Sigma_0 R_{thin}^2 R \times 
\int_0^{\infty} \frac{J_1(kR) \exp(-k|z|) k dk}{(1+k^2R_{thin}^2)^{3/2}}
\end{equation}
\begin{equation}
V_{disk}^2(R_{\odot},0) = 4 \pi G \Sigma_0 R_{thin} x^2 \times [I_0 (x) K_0(x) - I_1(x) K_1(x)]
\end{equation}
where $x = R_{\odot}/2R_{thin}$ (see Freeman 1970, Binney \& Tremaine 1987), and:
\begin{equation}
V_{bulge}^2(R,z) = R \frac{\partial \Phi_{bulge}}{\partial R} = \frac{G M_b R^2}{r(r+c)^2}
\end{equation}
\begin{equation}
V_{bulge}^2(R_{\odot},0) = \frac{G M_b R_{\odot}}{(R+c)^2}
\end{equation}

In what follows we will use eq. (11) to parametrize the dependence of $\overline{V_z^2}$ as a function 
of $R$, for a given $z$, where the term $\partial \overline{V_z^2} / \partial R$ is measured from
our data.

\subsection{Results}

Our theoretical $\overline{V_z^2}$ profiles are to be compared with data in one $z$ layer, namely
$ 1.5 < |z| < 2.0$ kpc, which includes $\sim 400$ stars, has the largest span in $R$, and has no
thin-disk contamination.
The expression for $\overline{V_z^2}$ contains several parameters which have a range of possible values. These parameters 
are: $\beta$ - ellipsoid alignment geometry, $z_{thick}$ - the scale height of the thick disk,
$R_{thick}$ - the scale length of the thick disk, $\partial \overline{V_z^2} / \partial R$ - the gradient of the
velocity dispersion,
$\Sigma_{thin}$ - the surface density of the
thin disk at the Sun's location, $V_c$ - the rotation of the LSR, $M_b$ - the bulge mass,
$c$ - the core radius of the bulge, $a$ the core radius of the halo, 
and finally $R_{thin}$ - the scale length of the thin disk. Parameter values are listed in Table 6
with the default value shown in bold.

The tests are made keeping all parameters fixed at default values and varying only one,
to isolate its impact on the profile. If its influence is minimal, we eliminate it as a potential 
uncertainty in the final determination of the thin-disk scale length. The default parameters were 
chosen as follows: the thick disk scale height and length are from De Jong et al. (2010),
the surface density of the thin disk at the Sun's location is from Holmberg \& Flynn (2004), the bulge 
parameters are from Johnston et al. (1995), the halo core radius is estimated as in Girard et al. (2006).
The value of $\partial \overline{V_z^2} / \partial R$ is from our measurements, while $R_{thin}$,
which we will eventually fit, is initially set to the most common value found in the literature based on starcounts.

\begin{table}
\caption{Model Parameters}
\begin{tabular}{lr}
\tableline
\\
\multicolumn{1}{c}{Parameter}  & \multicolumn{1}{c}{Values explored} \\
\tableline
 $\beta$ & 0, {\bf 1} \\
 $z_{thick}$ (kpc)  & 0.50, {\bf 0.75}, 1.00 \\
 $R_{thick}$ (kpc)  &  3.0, {\bf 4.1} \\
 $\partial \overline{V_z^2} / \partial R$ (km$^2$~s$^{-2}$~kpc$^{-1}) $ &  -686, {\bf -759}, -832 \\
 $\Sigma_{thin}$ ($M_{\odot}$~pc$^{-2}$) &  42.0, {\bf 56.0}  \\
 $V_c$ (km~s$^{-1}$) &  {\bf 220}, 250 \\
 $M_b$ ($10^{10} M_{\odot}$) &  1.0, {\bf 3.4}, 6.0 \\
 $c$  (kpc)          & 0.4,{\bf 0.7}, 1.0 \\
 $a$  (kpc)          &  5.0, {\bf 6.3}, 7.0 \\
 $R_{thin}$ (kpc)    & 1.5, 2.0, {\bf 2.5}, 3.0 \\
\tableline
\end{tabular}
\end{table}

In Figure 15, we show the effect  on the model profile of varying each parameter listed in Table 6, superposed on
the observations.  The default model is represented with a black line.
Clearly, large changes in $\beta$, $R_{thick}$, $\partial \overline{V_z^2} / \partial R$, $M_b$,
$c$, or $a$ do not affect the profile at the level that observations can discriminate. We are thus left 
with four parameters, of which $z_{thick}$, $\Sigma_{thin}$ and  $V_c$ more or less show a net vertical
shift in the profile when their values change, while $R_{thin}$ causes the profile to change its shape, becoming
steeper for shorter scale lengths. 
From the plots it is seen that the profiles are most sensitive to the parameters
$z_{thick}$ and $R_{thin}$.
Thus we apply a $\chi^2$ minimization for the fit of the observations with the model, to determine 
the most likely values of these two parameters.
We do so separately for the four combinations of the test values of the second-most relevant parameters $\Sigma_{thin}$ and $V_c$. 
The results are summarized in Table 7. Errors shown are formal errors derived from the fit, but are themselves highly uncertain due to the small number of bins.
Formally, the values of the fitted parameters vary within 0.2 kpc for both the scale height of the thick disk, and
the scale length of the thin disk.

We attempt to determine more realistic uncertainties in $z_{thick}$ and $R_{thin}$ by running Monte Carlo 
simulations of our binned data.
We generate a set of 1000 representations of the four $\sigma_{V_{z}}$ data points, using Gaussian deviates from the observed 
values and with dispersion equal to each bin's uncertainty.
Then, we run the $\chi^2$ minimization routine with fixed $\Sigma_{thin}$=56~$M_{\odot}$~pc$^{-2}$, and 
$V_c$=220 km~s$^{-1}$, and determine a set of 1000 determinations of $z_{thick}$ and $R_{thin}$. The resulting 1-$\sigma$ range 
is 0.1 kpc in $z_{thick}$, and 0.4 kpc in $R_{thin}$. With these uncertainties, the solutions in Table 7 are 
equally valid, and we can not distinguish between the two values for $\Sigma_{thin}$ and $V_c$. 
We therefore adopt as our best determination $z_{thick} = 0.7 \pm 0.1$ kpc, and $R_{thin} = 2.0 \pm 0.4$ kpc.
Our thick disk scale height is in very good agreement with the determination by
de Jong et al. (2010) of $0.75 \pm 0.07$ kpc from SDSS and with Girard et al. (2006) of 
$0.78 \pm 0.05$ kpc from 2MASS-selected red giants.

The thin disk scale length is also in reasonable agreement with starcount determinations which
range from roughly 2 to 3 kpc (Siegel et al. 2002, Cignoni et al. 2008 and references therein), with our value being 
on the low end of the range.

\begin{table}
\caption{Estimations of $z_{thick}$ and $R_{thin}$}
\begin{tabular}{llll}
\tableline
\\
 \multicolumn{1}{c}{Solution} & \multicolumn{1}{c}{$z_{thick}$} & \multicolumn{1}{c}{$R_{thin}$} & \multicolumn{1}{c}{$\chi^2$} \\
 & \multicolumn{1}{c}{(kpc)}  & \multicolumn{1}{c}{(kpc)} & \\
\tableline
 $\Sigma_{thin}$=42~$M_{\odot}$~pc$^{-2}$, $V_c$=220 km~s$^{-1}$ & $0.78\pm0.03$ & $1.86\pm0.12$ & 2.15 \\ 
 $\Sigma_{thin}$=56~$M_{\odot}$~pc$^{-2}$, $V_c$=220 km~s$^{-1}$ & $0.70\pm0.03$ & $1.97\pm0.12$ & 2.01 \\ \\
 $\Sigma_{thin}$=42~$M_{\odot}$~pc$^{-2}$, $V_c$=250 km~s$^{-1}$ & $0.66\pm0.03$ & $1.77\pm0.11$ & 2.24 \\ 
 $\Sigma_{thin}$=56~$M_{\odot}$~pc$^{-2}$, $V_c$=250 km~s$^{-1}$ & $0.60\pm0.02$ & $1.88\pm0.11$ & 2.08 \\ 
\tableline
\end{tabular}
\end{table}

\section{Formation of the Thick Disk}

Competing formation scenarios of the thick disk can be separated into two broad categories:
a) internal processes such as scattering/migration due to spiral arms (e.g., Sellwood \& Binney 2002, Roskar et al. 2008,
Loebman et al. 2010)
 and scattering by massive ($10^{8}$ M$_{\odot}$) clumps present in gas-rich, young galaxies (Bournaud et al. 2009), and
b) external processes such as accretion of many small satellites (Abadi et al. 2003), and
minor mergers with (Brook et al. 2004) and without (e.g., Villalobos et al. 2010, 
Villalobos \& Helmi 2008, Kazantzidis et al. 2009, Read et al. 2008) star formation during the merging process.
There may be some overlap between the two categories; for instance
mergers and perturbations from satellites can excite spiral structure and thus
induce radial migration (Quillen et al. 2009).

The more recent models of the second category include cosmologically motivated
merging histories of MW-type halos, which are important in assessing a realistic impact of
the merging process on the initial thin disk of the Galaxy (Read et al. 2008, Kazantzidis et al. 2009). 
However, not all models of the second category need a pre-existing thin disk.
For instance, in the Brook et al. (2004) model, the thick disk forms at an early epoch 
(between 10 and 8 Gyrs ago) characterized by
merging of gas-rich protogalaxies, while the thin disk forms in the 
following 8 Gyr, in a quiescent period. Thick disk stars are thus 
predominantly formed in situ, and only a
small fraction belongs to the satellites.
In the Abadi et al. (2003) model, the majority of stars in the thick disk are from accreted and disrupted satellites.
In the model by Villalobos \& Helmi (2008) (also Villalobos et al. 2010, Kazantzidis et al. 2009,
Read  et al. 2008), the thick disk is formed by the dynamical heating induced by 
a massive (5:1 mass ratio) satellite merging with a primordial thin disk.

It is not trivial to make a meaningful comparison between models, 
since each study is focused on a different aspect of this process:
likelihood of disk survival, structure and orbits of accreted satellites, 
structural parameters of the thick disk, various kinematical 
features of the thick disk, abundance patterns, etc.
However, a comparative study is presented by Sales et al. (2009) who use four models from the
literature and propose as a discriminator the distribution of the orbital eccentricity. The four 
models are radial migration (Roskar et al. 2008), accretion of many small satellites
(Abadi et al. 2003), dynamical heating due to a minor merger, without star formation, on a pre-existing disk (Villalobos \& Helmi 2008), and
gas-rich mergers during an active epoch, with in-situ star formation (Brook et al. 2004).

Here, we compare the eccentricity distribution of our observed thick-disk sample 
with those of the four models from Sales et al. (2009).
The observed distribution is obtained by integrating the orbits of stars in the 
Johnston et al. (1995) 
potential, which includes a bulge, a disk and a halo. We use only the stars 
with $ 1 < |z| < 3 $ kpc, and
$6 < R_{GC} < 9$ kpc. This is in order to match the samples in Sales et al. (2009) who used the normalized quantities
$ 1 < z/z_{0} < 3 $, and $ 2 < R_{GC}/R_d < 3$, where $z_0$ in the scale height of the thick disk, and
$R_d$ is the scale length of the thick disk. The scale height of the model thick disks
is $\sim 1$ kpc, except for the Abadi et al. (2003) model where it is 2.3 kpc. The scale lengths of the models
vary between 3 and 4 kpc (Tab. 1 in Sales et al. 2009).
Thus our sample may not be very well representative of the Abadi et al. (2003) model; our entire sample 
does not extend beyond $|z| = 3$ kpc (see Fig. 7).
The sample considered here for the eccentricity distribution consists of 1573 stars.

Eccentricities are calculated as $e = (r_a - r_p)/(r_a + r_p)$, where $r_a$ and $r_p$ are 
the apo- and pericenter distances. To account for errors in the observed quantities, we run
simulations with randomly generated proper motions, distances and radial velocities, as
drawn from the Gaussian distributions of their errors. For each star, a set of 50 realizations is made; 
then from the ensemble of realizations we construct the eccentricity distribution  shown in Figure 16
(shaded curve).
Each panel of Fig. 16 also shows the eccentricity distribution from one of the four models explored by Sales et al. (2009)
(black line). The observed distribution is quite asymmetric, with a peak at low eccentricities and
a long tail toward higher eccentricities in agreement with the Wilson et al. (2010) distribution 
also derived using RAVE data and with  the Dierickx et al. (2010) distribution derived using SDSS data.

From Fig. 16, the accretion model appears inconsistent with the observed distribution.
See also  Dierickx et al. (2010), for a sample at higher $|z|$ that better matches
the selection criteria of the model, but is still in disagreement with the resulting eccentricity distribution.
Their conclusion 
is similar to ours: there are not enough high eccentricity stars to satisfy the model distribution.
Similarly, the radial migration model is difficult to reconcile with the observations; 
there are not enough low eccentricity stars, and there 
are too many high eccentricity stars in the observations compared to the model. 

The two most favored models by our observations are the dynamical heating due to a minor merger, and 
the gas-rich merging.  Indeed, the major peak of the distribution and its long tail are well reproduced by 
both models.
Even if our observations do not have a secondary peak at very high eccentricities, 
as the heating model displays, this in not necessarily a strong disagreement. In the model, the secondary 
peak is due to stars from the merging satellite, and these preserve the initial orbit of the satellite.
A less eccentric initial orbit of the satellite may very well produce a distribution more in line with the 
observed one (see also Wilson et al. 2010 for a similar conclusion).

Another quantity to be used as a robust discriminator is the rotation velocity gradient with $z$, 
which here is found to be $-25.2 \pm 2.1$ km~s$^{-1}$ kpc$^{-1}$, and nearly -30 km~s$^{-1}$ kpc$^{-1}$ in
Girard et al. (2006), Chiba \& Beers (2000), Ivezi\'{c} et al. (2008). For instance, the radial migration model
predicts a shallower gradient of -17  km~s$^{-1}$ kpc$^{-1}$ (Loebman et al. 2010). 
A rather large gradient (consistent with the values presented here) is also predicted by the dynamical 
heating model of Villalobos \& Helmi (2008), provided
the satellite is on a low inclination ( $ < 30\arcdeg$) orbit. In fact, the dependence of
$\sigma_{V_R}$, $\sigma_{V_{\theta}}$  and $\sigma_{V_z}$ as a function of $R_{GC}$ seen in our sample
 are also consistent 
with the predictions of the low-inclination heating event described in Villalobos \& Helmi (2008): 
$\sigma_{V_R}$, $\sigma_{V_{\theta}}$ are nearly constant, while $\sigma_{V_z}$ decreases slowly with increasing $R_{GC}$
(see Fig. 13).
As a next step, it would be interesting to compare the predictions from the Brook et al. (2004) model
with our results for velocity  and velocity-dispersion gradients with $|z|$ and $R_{GC}$, to further help 
discriminate between models. 

Likewise, it would be helpful to investigate the predictions of the Brook et al. (2004)
model for the rotation velocity as a function of metallicity. Loebman et al. (2010) argue that the lack 
of correlation between these quantities (as observed by SDSS) favors the radial mixing model.
Our data span a small metallicity range, thus we are not able to reliably explore this issue.

In conclusion, our data favor the gas-rich merger model and the model describing a minor merger event 
heating a pre-existing disk for the origin of the thick disk.

\section{Summary}

We analyze the 3D kinematics of $\sim 4400$ red clump stars sampling the thin and thick disk of our Galaxy.
This sample is distributed mostly in quadrant four, above and below the plane, where there are no presently known overdensities
or other substructure. The sample probes a distance between 0.4 and 3 kpc
from the Galactic plane, and between 5 and 10 kpc from the GC, limits imposed by the magnitude selection of the
RAVE input catalog. For the subsample of stars that have metallicity estimates from RAVE, we infer that
the metallicity range of our red-clump sample is from -0.6 to +0.5 with a peak at $\sim -0.1$ dex.

We find that stars more distant than 0.7 kpc from the Sun have a net radial velocity of $12.7 \pm 3.1$  km~s$^{-1}$,
in the Galactic rest frame, while the vertical velocity is consistent with zero. Some recent studies using different tracers and 
probing different volumes in the Galaxy also show this offset. We argue that this motion is real, and we interpret it
as a mean motion of the LSR as defined by local samples, rather than an expansion of the Galaxy. 

After removing the contribution of the thin disk, we determine the $z$-gradient of the rotation velocity 
of the thick disk to be $\partial V_z / \partial z = -25.2 \pm 2.1$ km~s$^{-1}$ kpc$^{-1}$,
in reasonable agreement with previous determinations (Girard et al. 2006, Chiba \& Beers 2000).

The velocity dispersions at $|z|=1$ kpc are
$(\sigma_{V_R}, \sigma_{V_{\theta}}, \sigma_{V_z}) = (70.4, 48.0, 36.2) \pm(4.1,8.3,4.0)$ km~s$^{-1}$, and their
gradients are $ (\partial \sigma_{V_R}  / \partial z, \partial \sigma_{V_{\theta}}  / \partial z,\partial \sigma_{V_z}  / \partial z) =
(17.5, 17.1, 5.0) \pm (2.5, 5.0, 2.4)$ km~s$^{-1}$~kpc$^{-1}$ for $0.7 \le |z| \le 2.5$ kpc
 in agreement with
equilibrium models described in Girard et al. (2006).

The velocity dispersions $R_{GC}$-gradients are consistent with zero for the radial 
and rotational components, while for the vertical component, we obtain a gradient $ (\partial \sigma_{V_z} / \partial R) = -8 \pm 3$
km~s$^{-1}$~kpc$^{-1}$, for $6 \le R_{GC} \le 9$ kpc. This latter dependence is better described by the
Jeans equation, which allows us to also determine the scale length of the thin disk, and the scale height of the thick disk.
Our dynamical estimate of the thin disk scale length is $R_{thin} = 2.0 \pm 0.4$ kpc, and that of the
thick-disk scale height is $z_{thick} = 0.7 \pm 0.1$ kpc.

The tilt angles of the velocity ellipsoid are consistent with zero except for 
$\alpha_{Rz} = 8.6\arcdeg \pm 1.8\arcdeg$

By comparing the distribution of orbital eccentricities as determined from our data with those of the four models
explored by Sales et al. (2009), as well as from the inspection of other kinematical parameters,
we favor the gas-rich merging model and the minor merging heating of a pre-existing thin disk
 for the formation of the thick disk.

We acknowledge support by NSF grants AST04-0908996, AST04-07292 and AST04-07293. We are grateful to Laura Sales 
for providing the data for Figure 16, and we thank Laura Sales and Amina Helmi for helpful discussion. 

Funding for RAVE has been provided by: the Anglo-Australian Observatory; the Astrophysical Institute Potsdam; the Australian National University; the Australian Research Council; the French National Research Agency; the German Research foundation; the Instituto Nazionale di Astrofisica di Padova; the Johns Hopkins University; the National Science Foundation of the USA (AST09-08326); W.M. Keck foundation; the Macquarie University; the Netherlands Research School for Astronomy; the Natural Sciences and Engineering Research Council of Canada; the Slovenian Research Agency; the Swiss National Science Foundation; the Science \& Technology Facilities Council of the UK; Opticon; Strasbourg University; and the Universities of Groningen, Heidelberg and Sidney.

This publication makes use of data products from Two Micron All Sky Survey, which is a joint projects of the University of Massachusetts and the Infrared Processing and Analysis Center/
California Institute of Technology, funded by the NASA and NSF.

\clearpage

\begin{figure}
\includegraphics[scale=0.8]{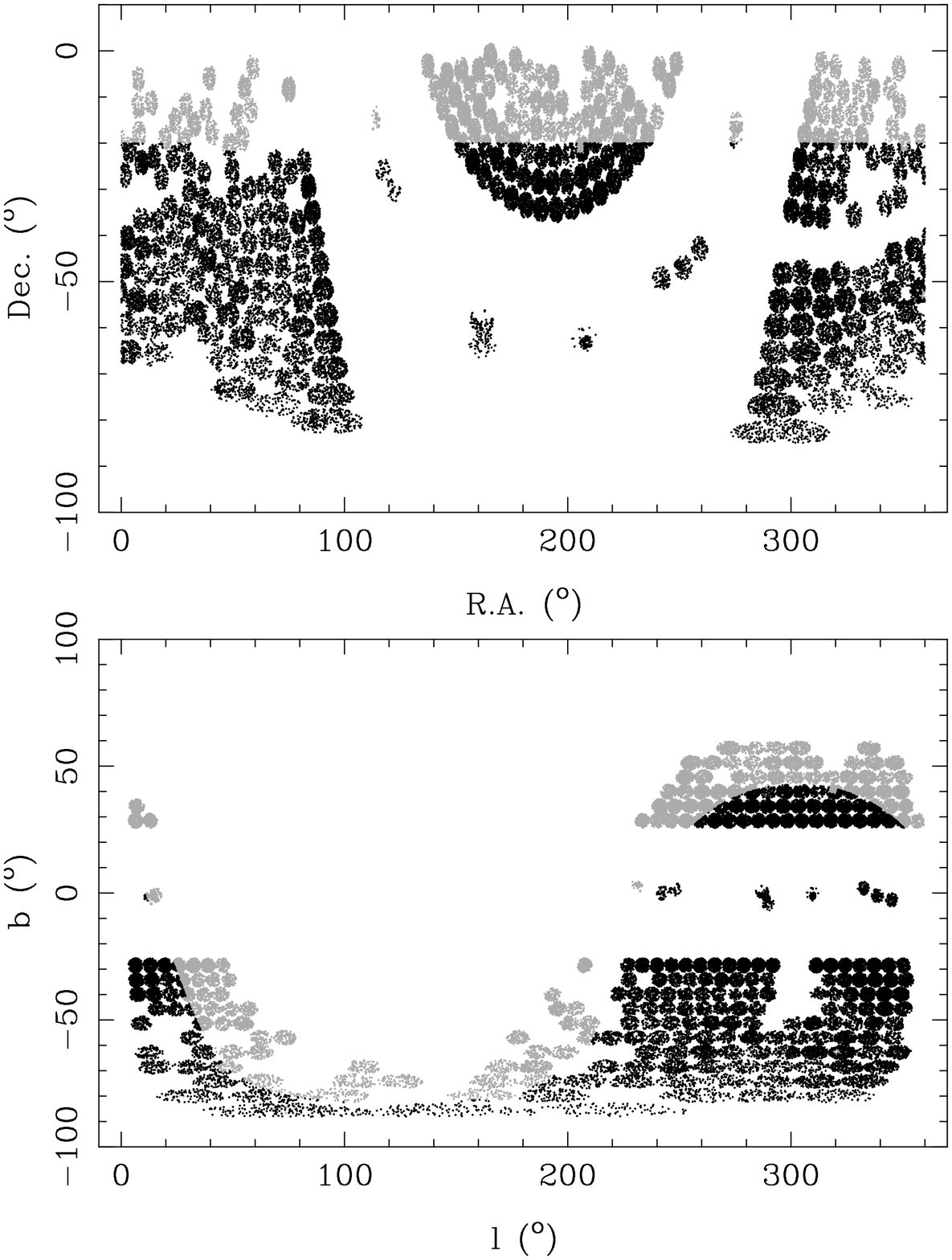}
\caption{Equatorial (top) and galactic (bottom) coordinates of the intersection between RAVE DR2 and
SPM4 data (black), and RAVE DR2 data (grey).}
\end{figure}

\begin{figure}
\includegraphics[scale=0.7]{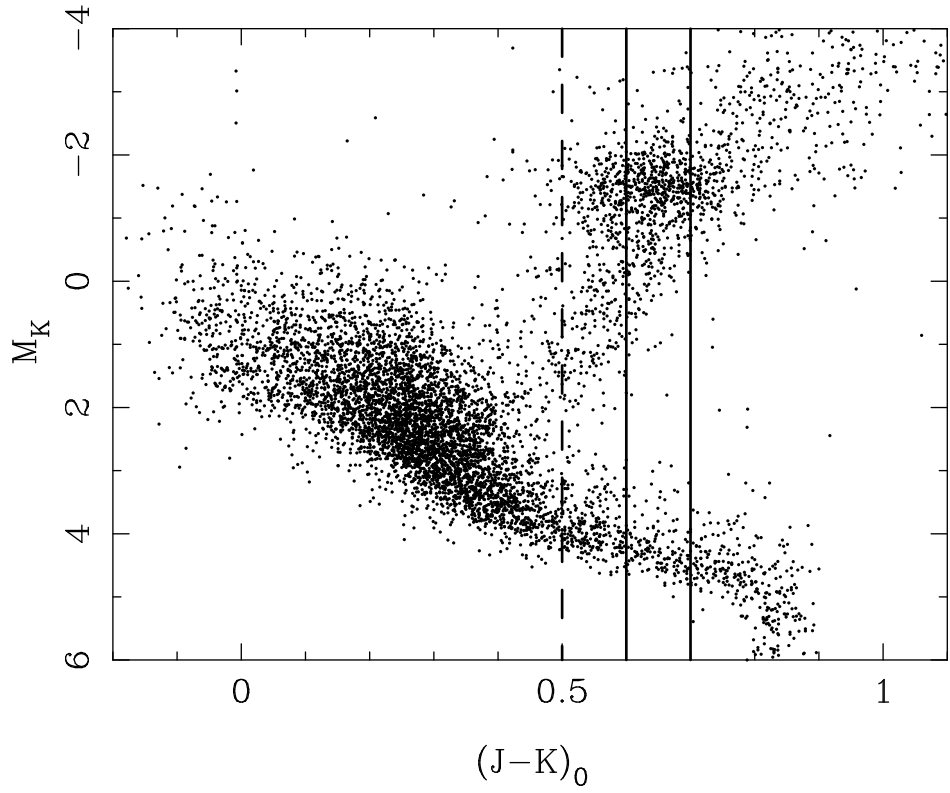}
\caption{$Hipparcos$ and 2MASS HR diagram for stars with parallax errors $\sigma_{\pi}/\pi < 0.15$.
The red-clump color selection often used is $0.5 \le (J-K)_0 \le 0.7$. In this study, we have used
$0.6 \le (J-K)_0 \le 0.7$.}
\end{figure}

\begin{figure}
\includegraphics[scale=0.7]{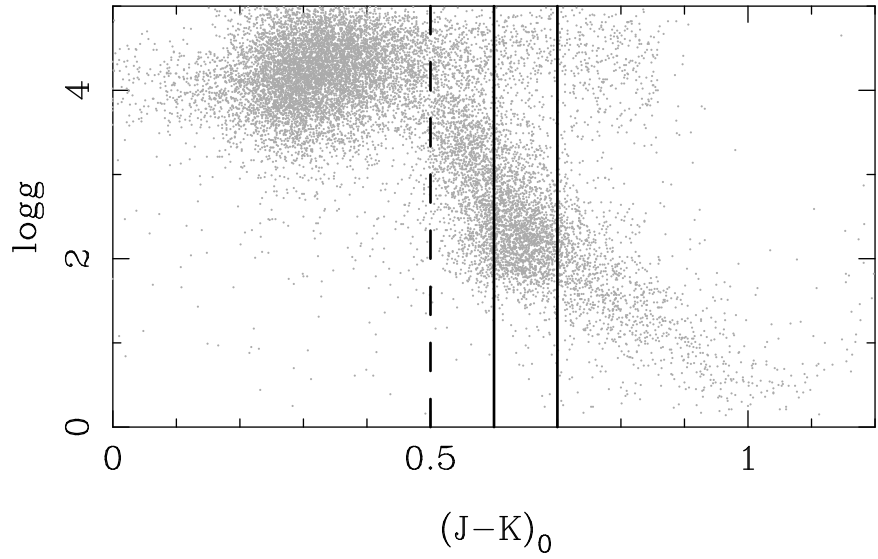}
\caption{Surface gravity as a function of color. The more restrictive color cut 
$0.6 \le (J-K)_0 \le 0.7$ avoids many of the subgiants where logg values fall in the range 3 to 4.}
\end{figure}

\begin{figure}
\includegraphics[scale=0.7]{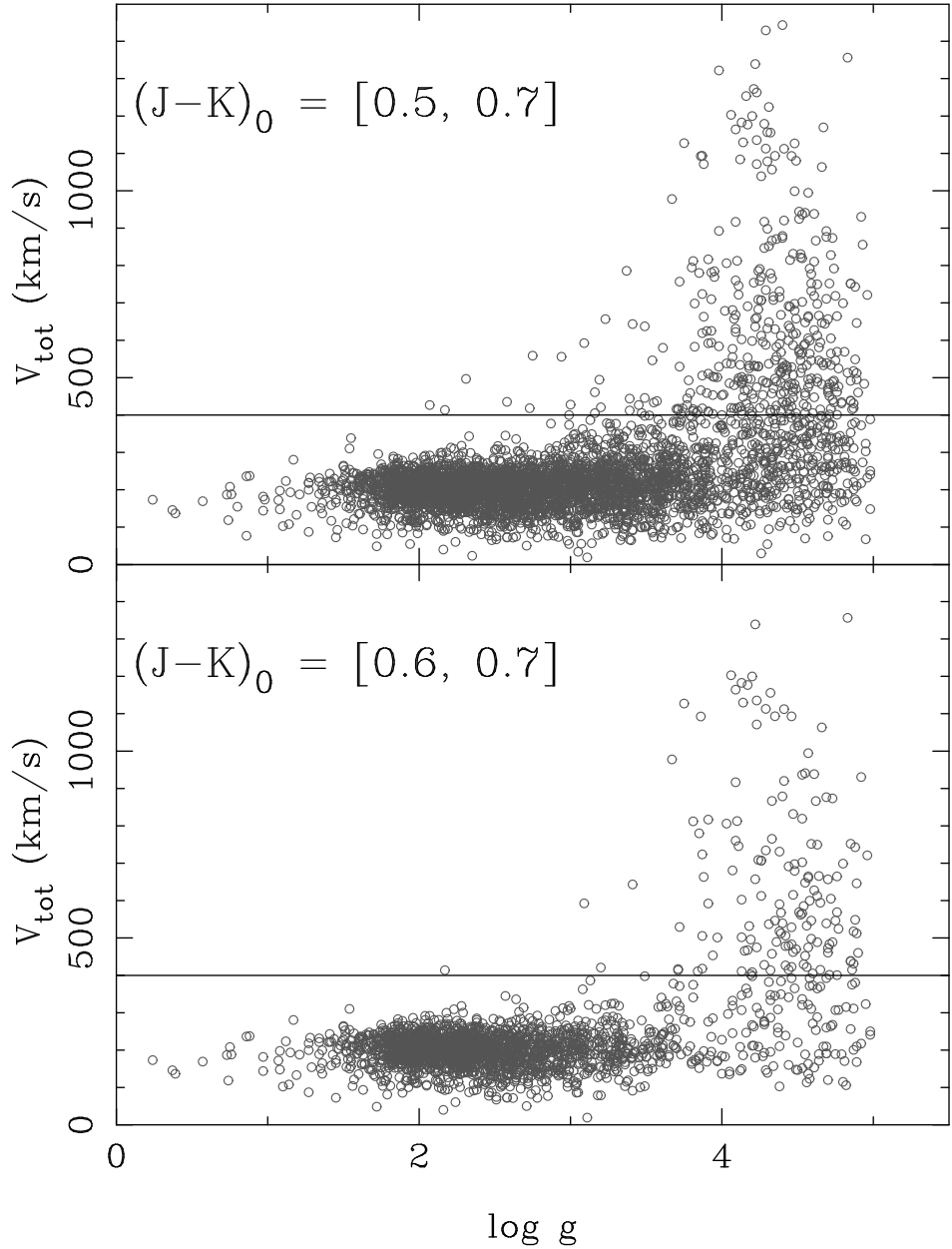}
\caption{Total velocity as a function of log~$g$ for two color ranges as specified in each panel.
The more restrictive color range avoids subgiant contamination. The horizontal line represents
the velocity cut for our red-clump sample.}
\end{figure}

\begin{figure}
\includegraphics[scale=0.7]{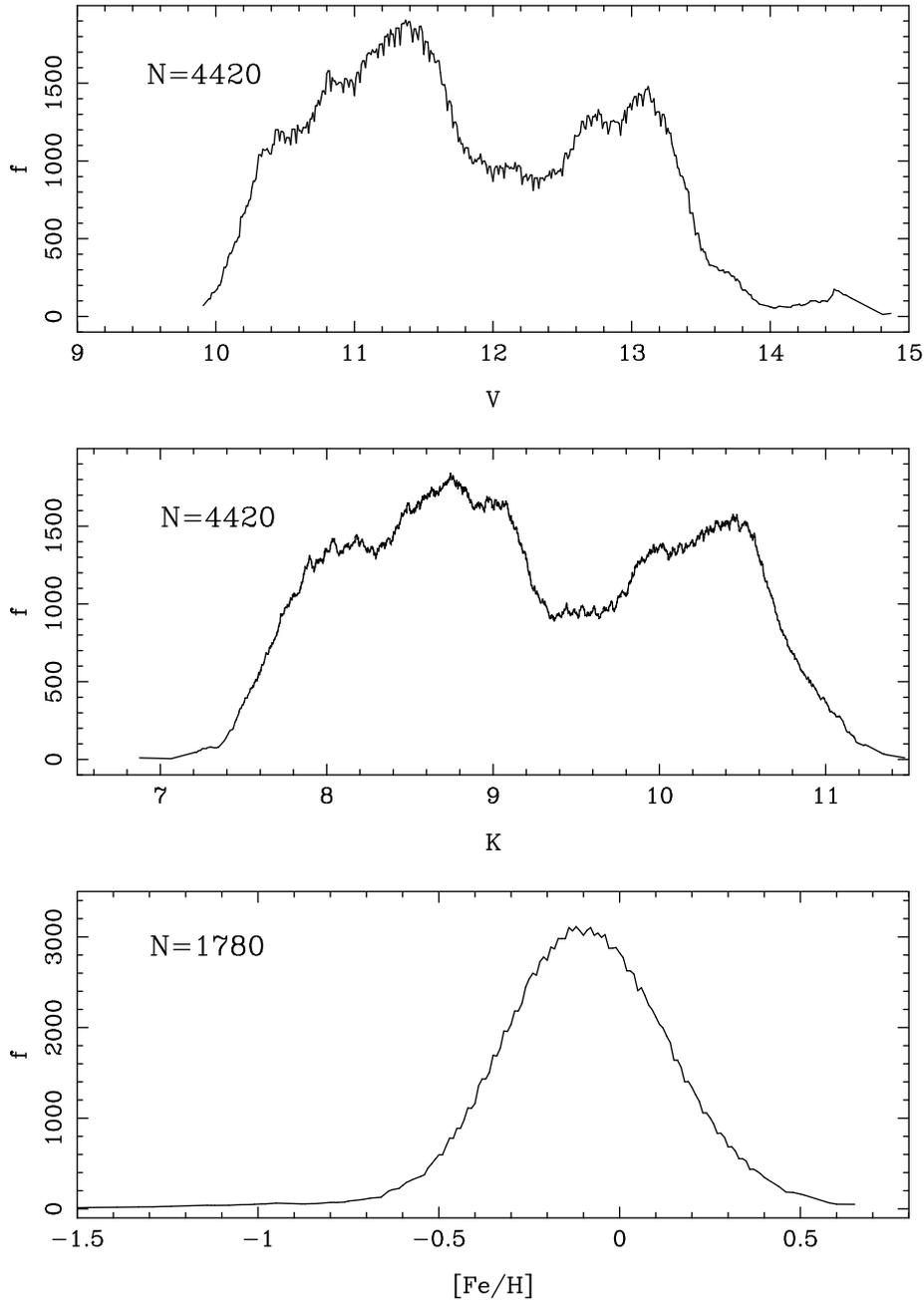}
\caption{Magnitude and metallicity distributions for our sample of red clump stars.}
\end{figure}

\begin{figure}
\includegraphics[scale=0.7]{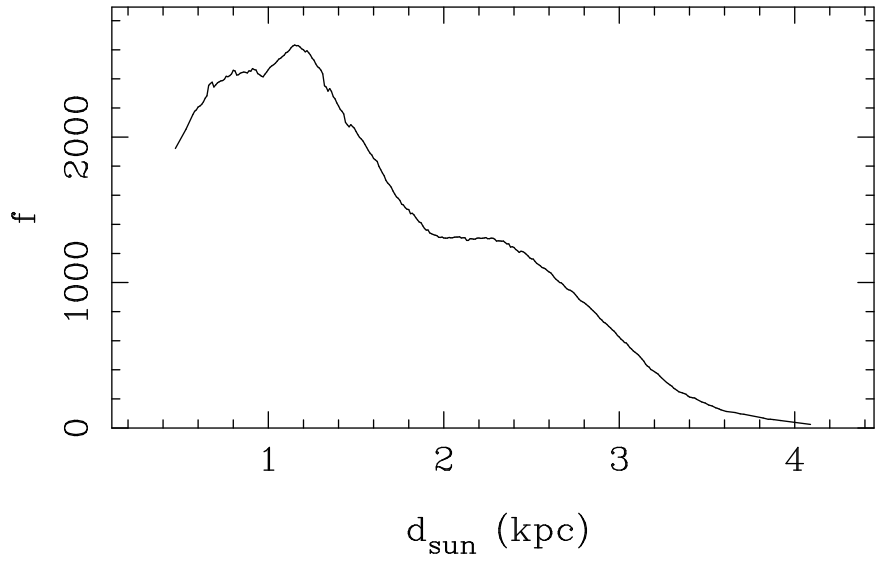}
\caption{Distribution of photometric distance from the Sun for the 4420 candidate red clump stars.}
\end{figure}

\begin{figure}
\includegraphics[scale=0.9]{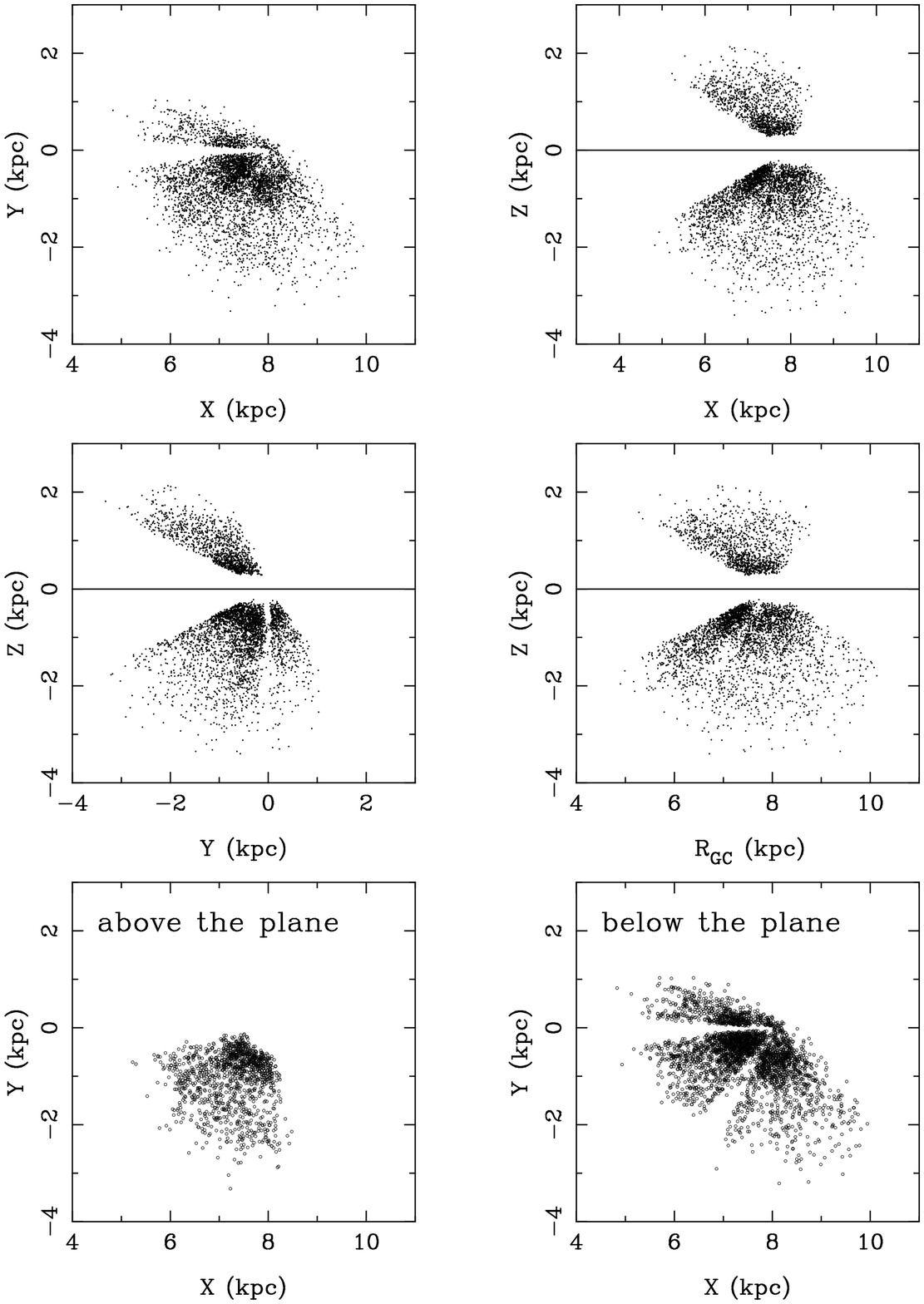}
\caption{Spatial distribution of the sample in various projections. The bottom panels show
the in-plane projections for above and below the plane samples.}
\end{figure}

\begin{figure}
\includegraphics[scale=0.8]{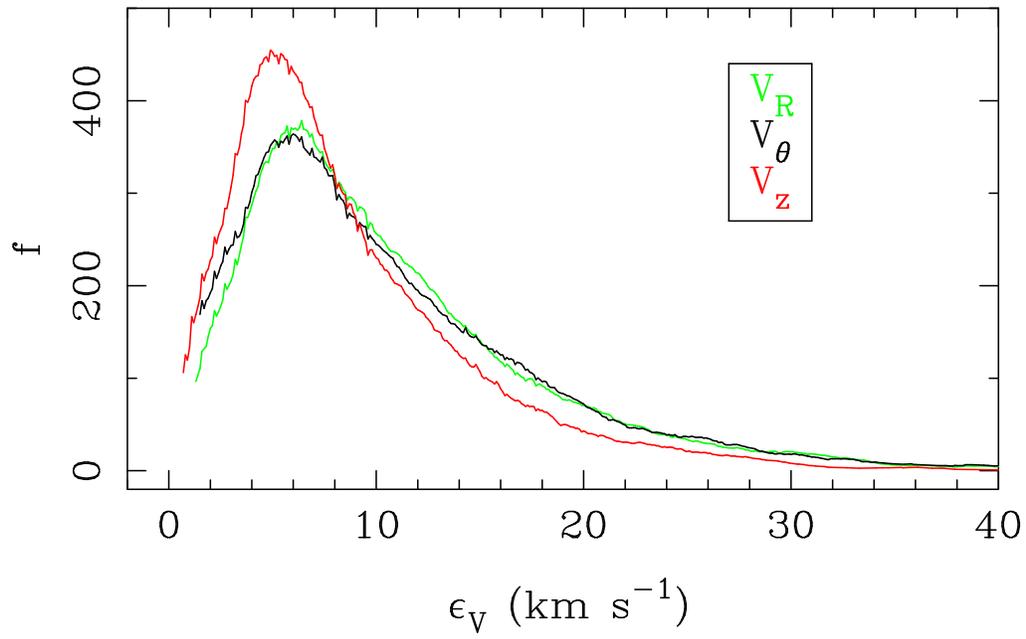}
\caption{Distribution of the estimated velocity errors in each component as labeled.}
\end{figure}

\begin{figure}
\includegraphics[scale=0.8]{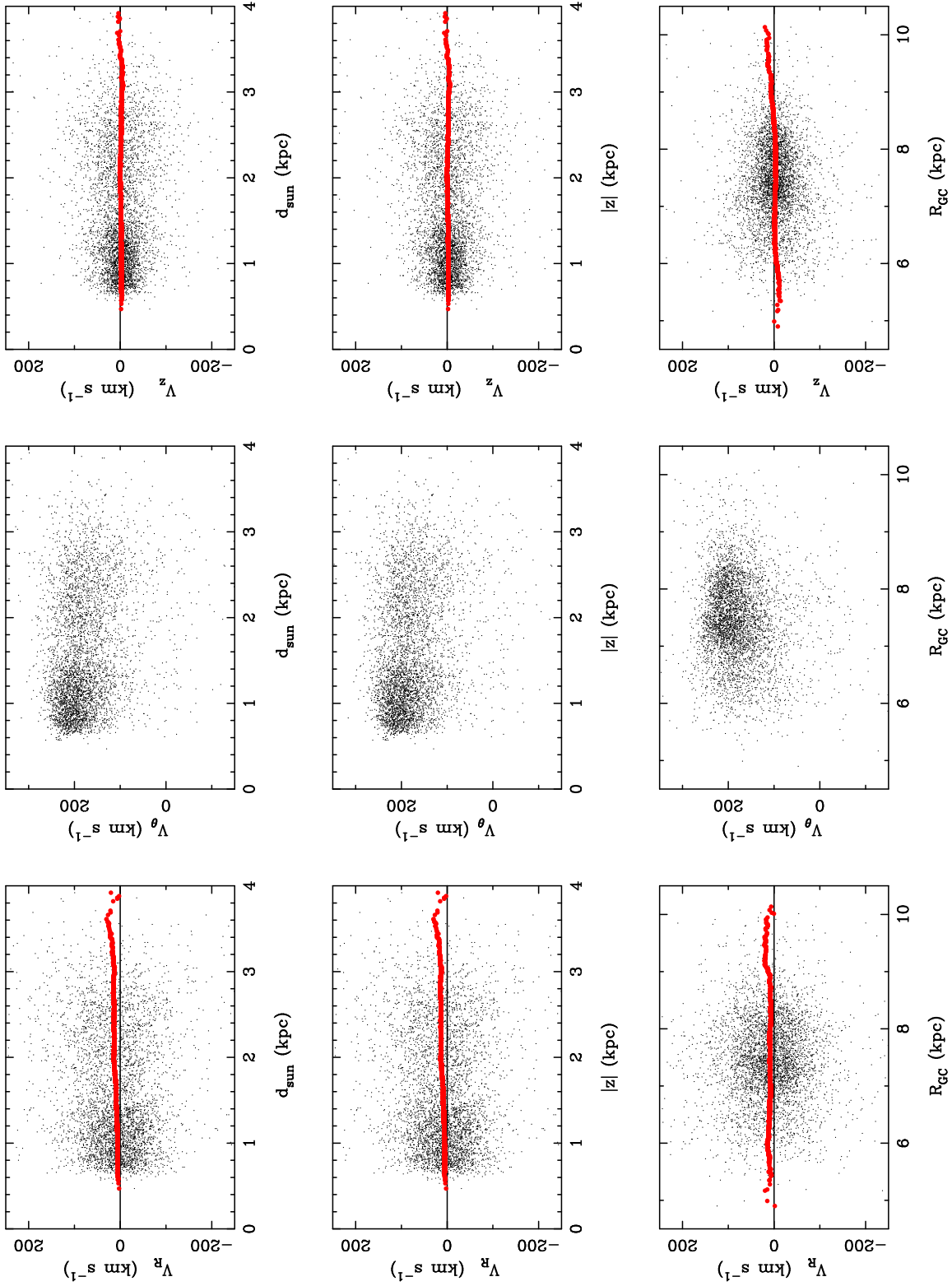}
\caption{Velocities in each component as a function of heliocentric distance (top), distance
from the plane (middle) and distance from GC (bottom) for the entire sample.
For $V_R$ and $V_z$, we also show a moving mean (red).}
\end{figure}

\begin{figure}
\includegraphics[scale=0.9]{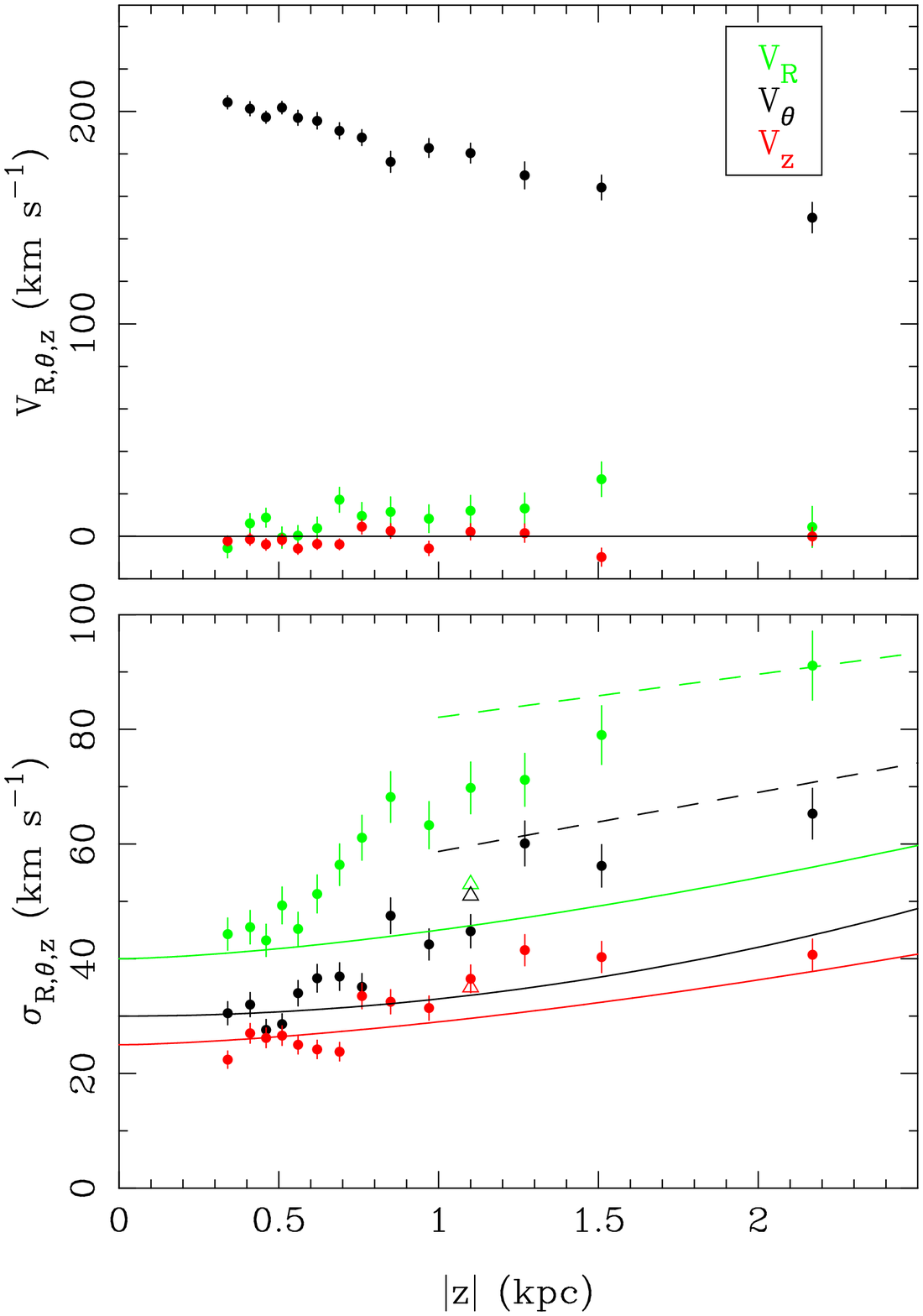}
\caption{Average velocities (top) and velocity dispersions (bottom) as a function of distance from the plane for
the entire sample.
Each velocity component is represented with a given color as labeled. 
Solid lines show the B10 dependence, while dashed lines show the Girard et al. (2006) dependence. 
The triangles represent the C10 values. The velocity dispersions are intrinsic, i.e. have been corrected for
measuring errors, but have not yet been corrected for thin-disk contamination.}
\end{figure}

\begin{figure}
\includegraphics[scale=0.9]{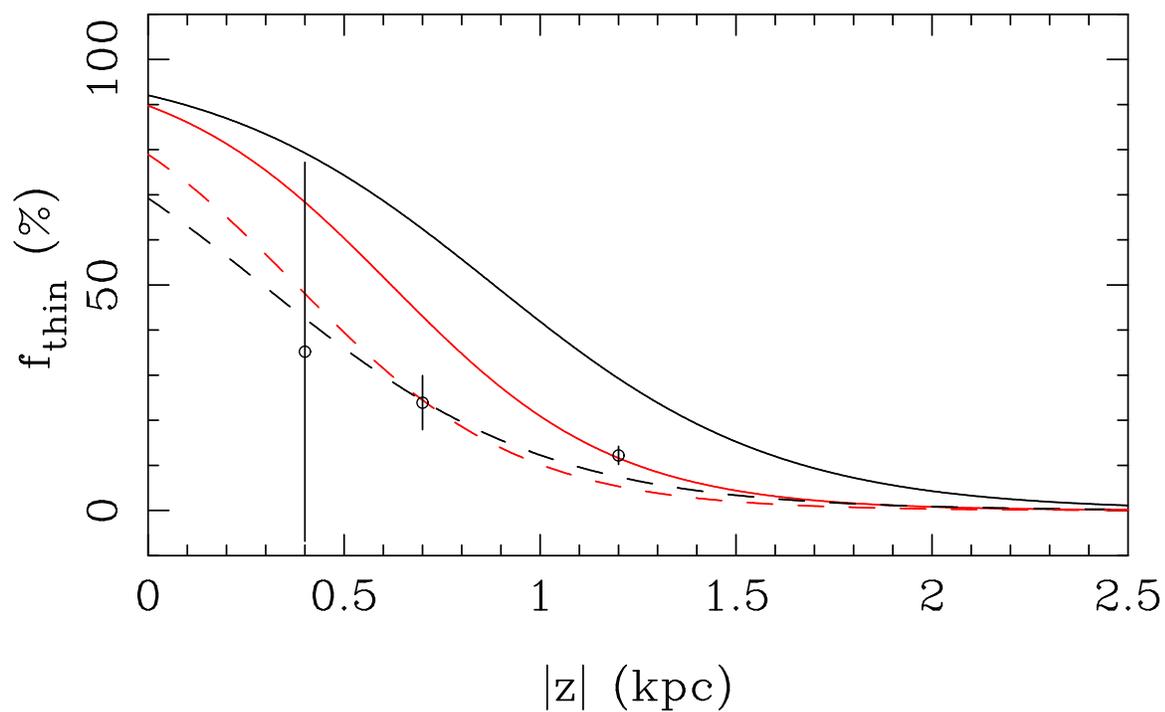}
\caption{Fraction of thin disk stars as a function of distance from the Galactic plane.
Solid lines represent two density laws from C05, while dashed lines represent the same laws normalized
to our measured fraction of thin-to-thick disk stars at $\overline{z} = 0.7$ kpc. The point with error bars are the 
results from Table 1: our determination of the thin-disk contamination based on dividing our sample in 
three $z$-distance bins.}
\end{figure}

\begin{figure}
\includegraphics[scale=0.8]{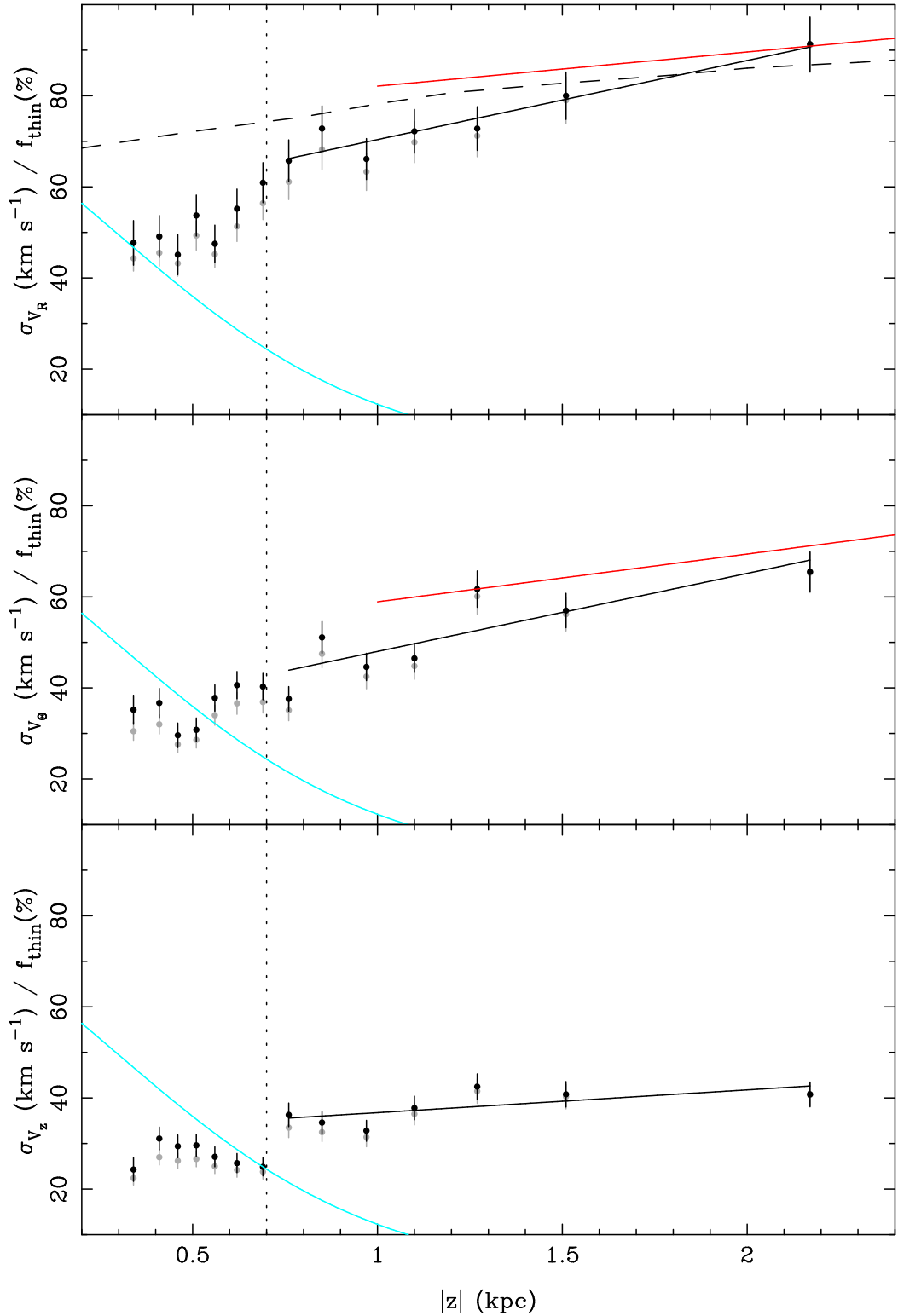}
\caption{Intrinsic velocity dispersions and thin-disk fraction as a function of distance from the plane.
Dispersions uncorrected for think-disk contamination 
are shown in grey, while those with the correction are shown in black.
The fraction of thin disk stars is shown with a blue line. The red lines show the gradients of the dispersions as
determined by Girard et al. (2006). Our linear fits are shown with a black line. 
The dashed line in the top panel shows one of the theoretical equilibrium models calculated in Girard et al. (2006) that fit their
data well.
The dotted line marks the
limit of $|z| = 0.7$ kpc, above which our fits linear are made.}
\end{figure}

\begin{figure}
\includegraphics[scale=0.8]{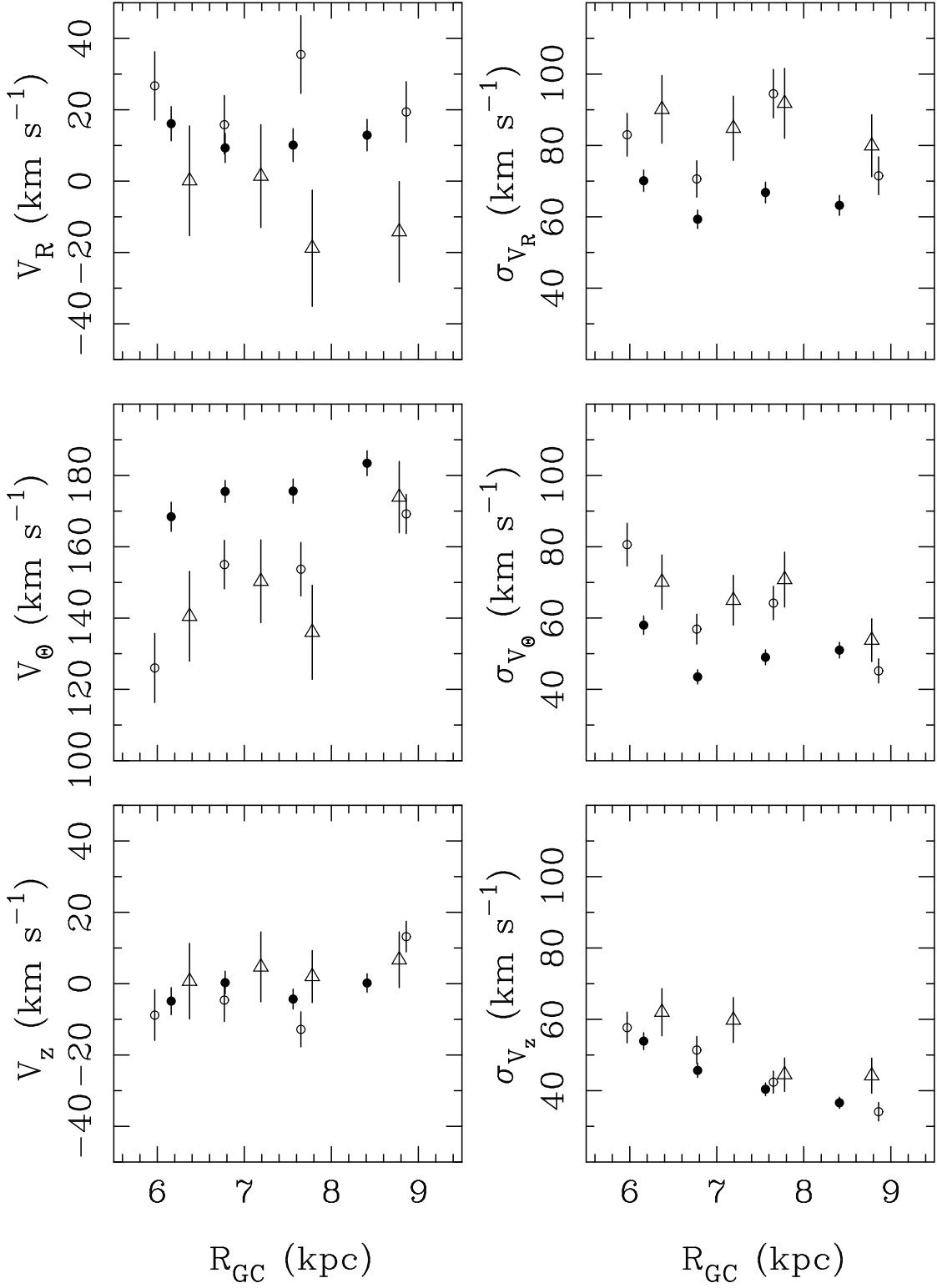}
\caption{Velocity (left) and velocity-dispersion (right) as a function  of $R_{GC}$.  Each row corresponds to 
one velocity component as labeled. Also, each panel includes data for three $z$-layers:
$1.0 < |z| < 1.5$ (filled circles), $1.5 < |z| < 2.0$ (open circles), and $2.0 < |z| < 2.5$ (open triangles).}
\end{figure}

\begin{figure}
\includegraphics[scale=0.8]{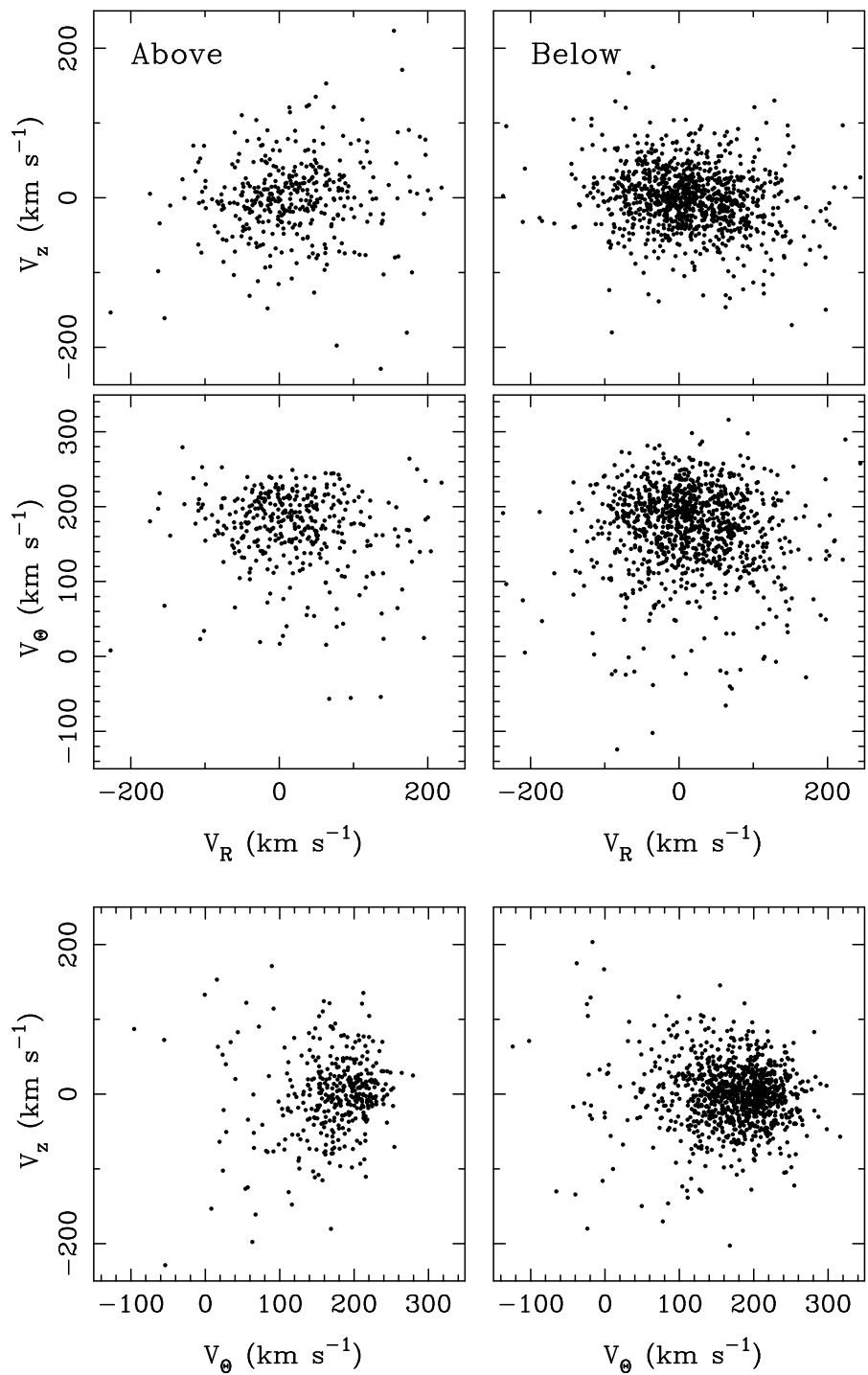}
\caption{Velocity pairs for the above- (left) and below-the-plane (right) samples. Stars are 
selected to be within $ 1 < |z| < 2 $ kpc, and $ 7 < R_{GC} < 9 $ kpc. There are 227 (538) stars in the
above(below)-the-plane samples.}
\end{figure}

\begin{figure}
\includegraphics[scale=0.85]{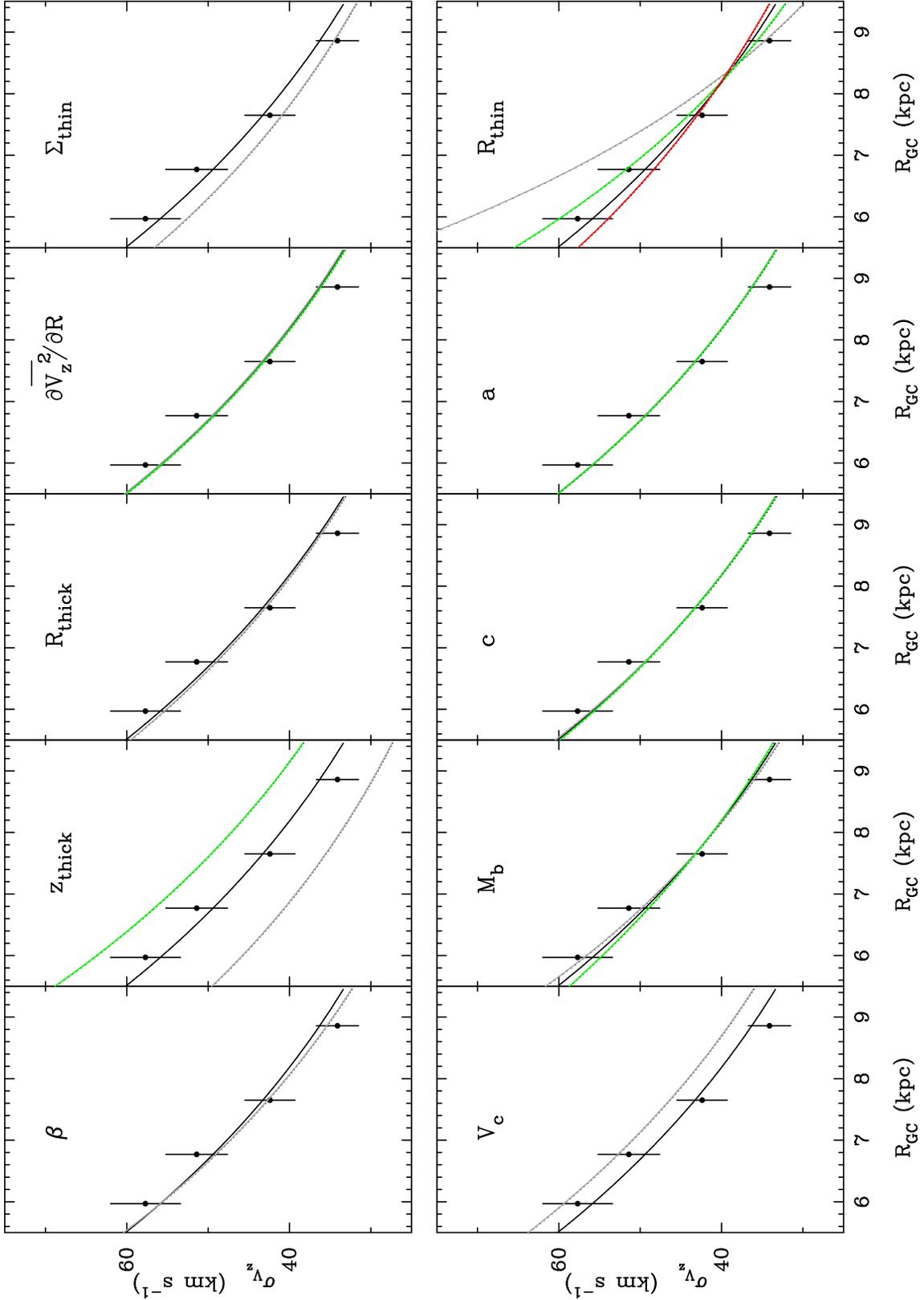}
\caption{$V_z$ dispersion as a function of $R_{GC}$ for various models (lines) and
for observations in the layer $ 1.5 < |z| < 2.0$ kpc (black circles). Each panel shows variation in one 
parameter as labeled. The default model is shown with a black line. For $z_{thick}$, the largest scaleheight (1.0 kpc) is shown with a green line. For $R_{thin}$, the largest 
value (3 kpc) is shown with a red line, and is the least steep profile.}
\end{figure}

\begin{figure}
\includegraphics[scale=0.7,angle=-90]{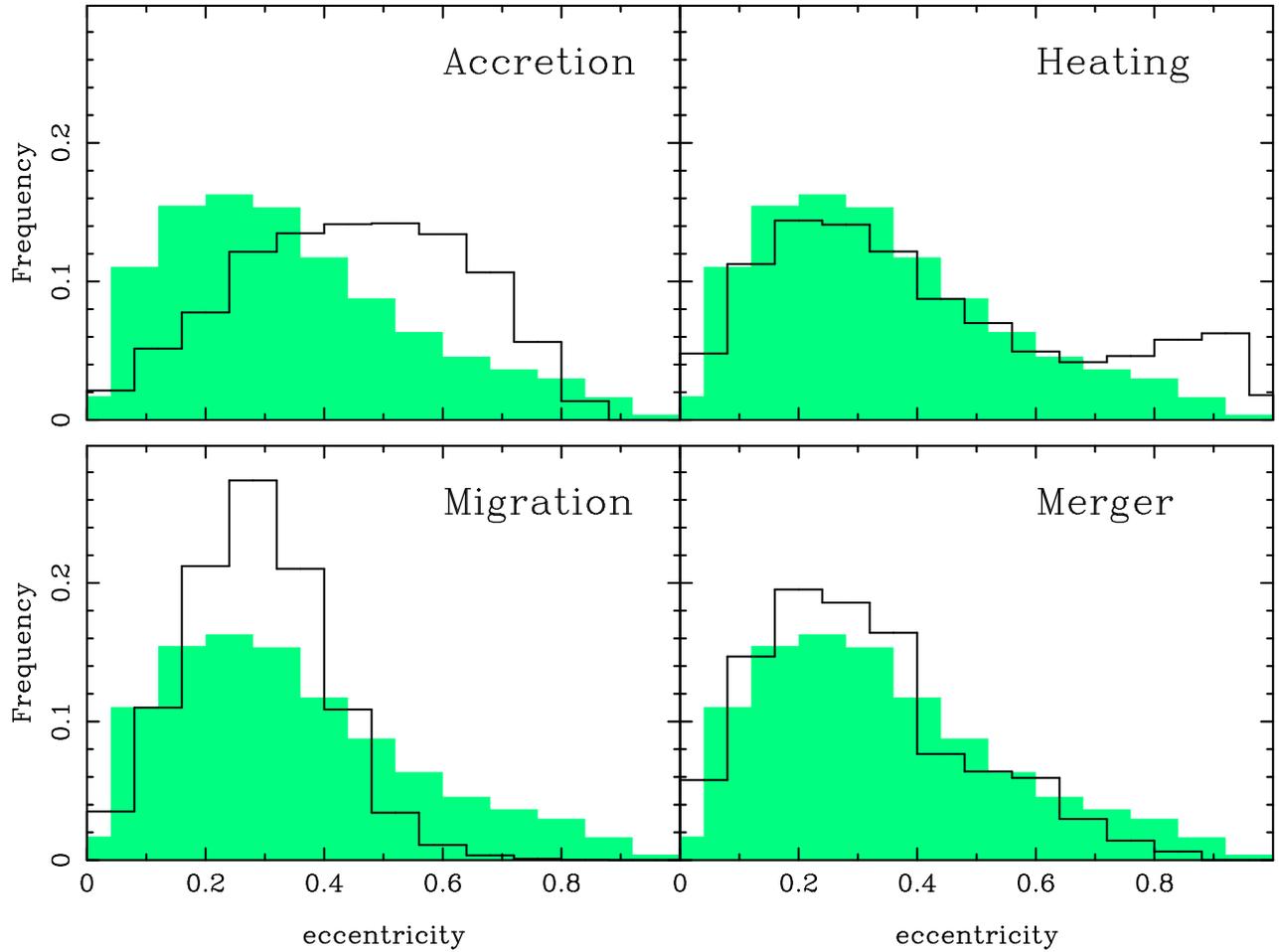}
\caption{Eccentricity distributions as determined from our sample of red clump stars (shaded) and from
the four models explored by Sales et al. (2009) (black line). The observed distribution includes 1573 star
within $1 < |z| < 3$ and $6 < R_{GC} < 9$.} 
\end{figure}

\end{document}